\documentclass[prb,twoside,twocolumn,10pt,floatfix,showpacs,citeautoscript,]{revtex4-2}

\usepackage{amsmath}
\usepackage{amssymb}
\usepackage{booktabs}
\usepackage{dcolumn}
\usepackage{glossaries}
\usepackage{graphicx}
\usepackage[version=4]{mhchem}
\usepackage{siunitx}
\usepackage{tabularx}
\usepackage[dvipsnames]{xcolor}

\usepackage{xpatch}
\makeatletter
\xpatchcmd{\@ssect@ltx}{\@xsect}{\protected@edef\@currentlabelname{#8}\@xsect}{}{}% Patch \<section>*
\xpatchcmd{\@sect@ltx}{\@xsect}{\protected@edef\@currentlabelname{#8}\@xsect}{}{}% Patch \<section>
\makeatother

\usepackage[colorlinks=true]{hyperref}
\hypersetup{
    pdffitwindow=false,
    pdfstartview={FitH},
    pdfnewwindow=true,
    colorlinks=true,
    linkcolor=NavyBlue,
    citecolor=NavyBlue,
    filecolor=NavyBlue,
    urlcolor=NavyBlue,
}

\hypersetup{
    pdfauthor={Erik Fransson, Julia Wiktor, and Paul Erhart},
    pdftitle={Phase transitions in inorganic halide perovskites from machine learning potentials},
}

\graphicspath{{figures/}}

\newcolumntype{d}{D{.}{.}{-1}}

\setacronymstyle{long-short}
\newacronym{cx}{vdW-DF-cx}{van-der-Waals-density functional with consistent exchange}
\newacronym{dft}{DFT}{density functional theory}
\newacronym{gpu}{GPU}{graphical processing unit}
\newacronym{md}{MD}{molecular dynamics}
\newacronym{mlp}{MLP}{machine learning potential}
\newacronym{nep}{NEP}{neuroevolution potential}
\newacronym{pes}{PES}{potential energy surface}
\newacronym{rmse}{RMSE}{root-mean-square error}
\newacronym{scan}{SCAN}{strongly constrained and appropriately normed}
\newacronym{soap}{SOAP}{smooth overlap of atomic positions}
\newacronym{xc}{XC}{exchange-correlation}

\DeclareSIUnit\angstrom{\text{Å}}
\DeclareSIUnit\atom{\text{atom}}
\DeclareSIUnit\step{\text{step}}
\renewcommand{\vec}[1]{\ensuremath\boldsymbol{#1}}
\renewcommand{\epsilon}[0]{\varepsilon}

\global\let\oldnewlabel\newlabel
\gdef\newlabel#1#2{\newlabelxx{#1}#2}
\gdef\newlabelxx#1#2#3#4#5#6{\oldnewlabel{#1}{{#2}{#3}}}
\let\newlabel\oldnewlabel
\newcommand{\addchalmers}{
    Department of Physics,
    Chalmers University of Technology,
    SE-41296, Gothenburg, Sweden
}

\begin{document}

\title{
    Phase transitions in inorganic halide perovskites from machine learning potentials
}

\author{Erik Fransson}
\author{Julia Wiktor}
\author{Paul Erhart}
\affiliation{\addchalmers}
\email{erhart@chalmers.se}

\begin{abstract}
The atomic scale dynamics of halide perovskites have a direct impact not only on their thermal stability but their optoelectronic properties.
Progress in machine learned potentials has only recently enabled modeling the finite temperature behavior of these material using fully atomistic methods with near first-principles accuracy.
Here, we systematically analyze the impact of heating and cooling rate, simulation size, model uncertainty, and the role of the underlying exchange-correlation functional on the phase behavior of \ce{CsPbX3} with X=Cl, Br, and I, including both the perovskite and the $\delta$-phases.
We show that rates below approximately \SI{30}{\kelvin\per\nano\second} and system sizes of at least a few ten thousand atoms are indicated to achieve convergence with regard to these parameters.
By controlling these factors and constructing models that are specific for different exchange-correlation functionals we then show that the semi-local functionals considered in this work (SCAN, vdW-DF-cx, PBEsol, and PBE) systematically underestimate the transition temperatures separating the perovskite phases while overestimating the lattice parameters.
Among the considered functionals the vdW-DF-cx functional yields the closest agreement with experiment, followed by SCAN, PBEsol, and PBE.
Our work provides guidelines for the systematic analysis of dynamics and phase transitions in inorganic halide perovskites and similar systems.
It also serves as a benchmark for the further development of machine-learned potentials as well as exchange-correlation functionals.
\end{abstract}

\maketitle

\section*{Introduction}

Halide perovskites are among the most intensively studied materials of the last decade due to their attractive properties for applications in, for example, solar energy harvesting and lighting \cite{kojima2009organometal,kim2012lead,hodes2013perovskite,van2018recent}.
Similar to their oxide counterparts many of these materials exhibit several different phases that are connected through soft modes and continuous or weak first-order phase transitions \cite{tyson2017large,liu2022neutron}.
This complex dynamic behavior turns out to be intimately connected to their remarkable optoelectronic properties.

In this context, electronic structure calculations play a crucial role as they can provide detailed insight into the atomic scale dynamics and microscopic coupling mechanisms.
Static calculations can, however, only provide limited information due the strong anharmonicity associated with the soft modes \cite{yaffe2017local, marronnier2017structural, marronnier2018anharmonicity, bechtel2019finite}.
This has motivated a number of dynamic studies based on \textit{ab-initio} \gls{md} simulations \cite{carignano2015thermal, wiktor2017predictive, BokLahRam17, mladenovic2018effects, zhu2022probing, gebhardt2022electronic, girdzis2020revealing, cannelli2022atomic} and, more recently, \glspl{mlp} \cite{jinnouchi2019phase, lahnsteiner2019long, thomas2019machine, zhou2020structural, mangan2021dependence, bokdam2021exploring, gruninger2021microscopic, LahBok22, BraGoeVan22, FraRosEri22}.
From such simulations one can then obtain, for example, transition temperatures \cite{jinnouchi2019phase} or structural information at finite temperatures \cite{carignano2015thermal, ghosh2017good, cannelli2022atomic}.

The accuracy of such simulations is, however, limited by several factors, most notably (1) sampling time, (2) system size, (3) the quality of the \gls{xc} functional, and, in the case of \glspl{mlp}, (4) the model uncertainty.
Sampling time and system size are in particular a problem for \textit{ab-initio} \gls{md} simulations, which are typically limited to time scales of a few ten picoseconds and system sizes on the order of a \num{1000} atoms.
Previous \gls{mlp} based studies were able to extend these ranges to total run times of a few nanoseconds using about a thousand atoms \cite{jinnouchi2019phase, lahnsteiner2019long, LahBok22}.

Here, we carry out a systematic analysis of the four factors described above.
We consider \ce{CsPbX3} with X=Cl, Br, and I and the \gls{scan} \cite{SunRuzPer15}, \gls{cx} \cite{BerHyl2014}, PBEsol \cite{PerRuz2008}, and PBE \cite{PerBurErn} \gls{xc} functionals.
The \gls{pes} is mapped using third-generation \glspl{nep} models and sampled using the \textsc{gpumd} package.
The latter provides an efficient \gls{nep} implementation that enables us to routinely sample systems comprising on the order of \num{60000} atoms for 100 or more nanoseconds.

We show that well converged results can be achieved using systems containing several ten thousand atoms and heating/cooling rates on the order of \SI{30}{\kelvin\per\nano\second} or lower.
%Maybe like this??:
%To assess the uncertainty of our \gls{nep}, we construct and compare ensambles of models trained on different splits of the training data. When such ensambles lead to significantly differing results, we bootstrap the models by creating new structures via \gs{md} runs and add them to the training set. Such a procedure leads to gls{nep} with an uncertainty that is comparable or lower than the training errors.
Using bootstrapping and ensembles of models we are able to readily generate accurate \gls{nep} models with an uncertainty that is comparable or lower than the training errors.

By controlling rate and size effects as well as model errors, we are able to isolate the impact of the underlying \gls{xc} functionals and thus to assess quantitatively the quality of different \gls{xc} functionals for the description of phase transitions and finite temperature properties of halide perovskites.
We find the \gls{cx} functional to perform the best among the \gls{xc} functionals considered here when comparing transition temperatures and lattice constants to experimental data.

In the following section we analyze in order rate and size effects (Sect.~\nameref{sect:rate-and-size-effects}), mode uncertainty (Sect.~\nameref{sect:model-uncertainty}), the impact of the \gls{xc} functional (Sect.~\nameref{sect:xc-functional}), and finally the transition temperature between the $\delta$ and perovskite phases (Sect.~\nameref{sect:delta-phase}).
We then summarize and discuss the outcome of this analysis (Sect.~\nameref{sect:discussion}).

\section*{Results}

\subsection*{Rate and size effects}
\label{sect:rate-and-size-effects}

The different perovskite phases are structurally closely related and connected through phase transitions with mixed continuous-first order character \cite{Hirotsu1974, Rodov2003, Malyshkin2020, Klarbring2019}.
For the \ce{CsPbX3} (with X=Cl, Br, or I) materials considered in this study the perovskite lattice transforms with increasing temperature from an orthorhombic phase ($Pnma$) via a tetragonal phase ($P4/mbm$) to a cubic phase ($Pm\bar{3}m$).
Since these transitions do not involve a switch in the sign of the Glazer angles between the orthorhombic ($a^-a^-c^+$) and tetragonal phases ($a^0a^0c^+$), unlike, e.g., \ce{MAPbI3} \cite{jinnouchi2019phase}, it is possible to observe these transitions in \gls{md} simulations.
Due to the remaining first-order character and the extreme heating/cooling rates that can be realized in \gls{md} simulations, one can, however, nonetheless anticipate some degree of hysteresis.

\begin{figure}[bt]
\centering
\includegraphics{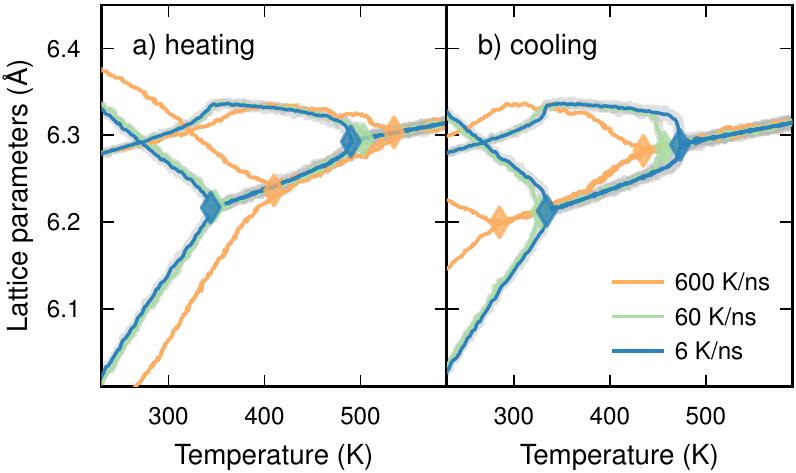}
\includegraphics{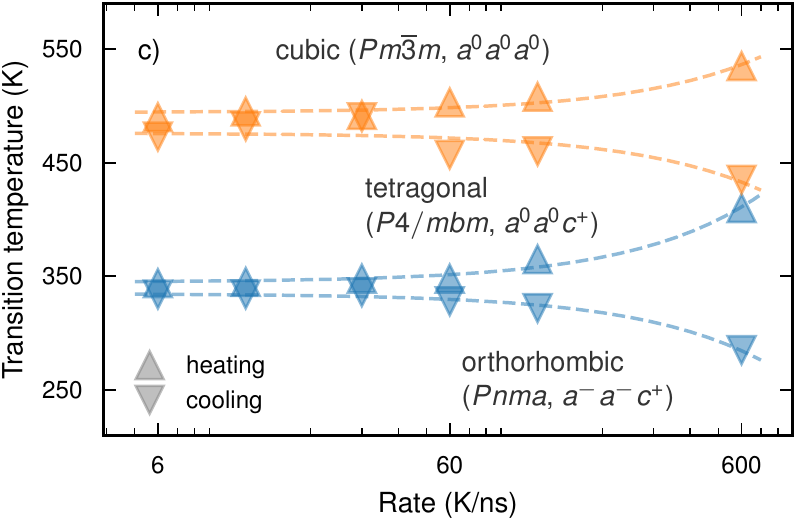}
\caption{
    \textbf{Convergence with heating/cooling rate.}
    Lattice parameters for \ce{CsPbI3} as a function of temperature from \gls{md} simulations with varying (a) heating and (b) cooling rates for supercells comprising \num{61440} atoms.
    The transition temperatures extracted from these data are indicated by diamonds.
    The gray lines show the raw data whereas the colored lines show the data after application of a Hamming window of \SI{0.6}{\kelvin} (see \autoref{sfig:hamming-window-lattice-parameters}).
    (c) Transition temperatures as a function of heating/cooling rate.
    All results were obtained using the full model (Sect.~\nameref{sect:model-construction}) based on the \gls{cx} exchange-correlation functional.
}
\label{fig:rate-effects}
\end{figure}

A further aggravating factor is the finite system size.
For small supercells the fluctuations are naturally larger, which renders it more challenging to achieve converged results.
In this section, we discriminate the effects of heating/cooling rate and system size by considering in detail \gls{md} simulations for \ce{CsPbI3} using the full model based on the \gls{cx} functional (Sect.~\autoref{sect:model-construction}).

\subsubsection*{Rate effects}
\label{sect:rate-effects}

To separate rate from size effects, we first consider the former in the large size limit, using a supercell comprising \num{61440} atoms, equivalent to 16x16x12 primitive orthorhombic (20-atom) unit cells.

On heating all simulations yield the correct (experimentally observed) sequence of phases irrespective of heating rate.
On cooling, this sequence is reversed again regardless of rate.
At the cubic-to-tetragonal transition, for a small number of simulations, one can, however, observe the simultaneous formation of multiple tetragonal domains with incompatible orientations, which can lead to the formation of domain boundaries.
Since the moving of these boundaries involves a nucleation-and-growth mechanism, they remain in the simulated structure upon cooling.

\begin{figure}
\centering
\includegraphics{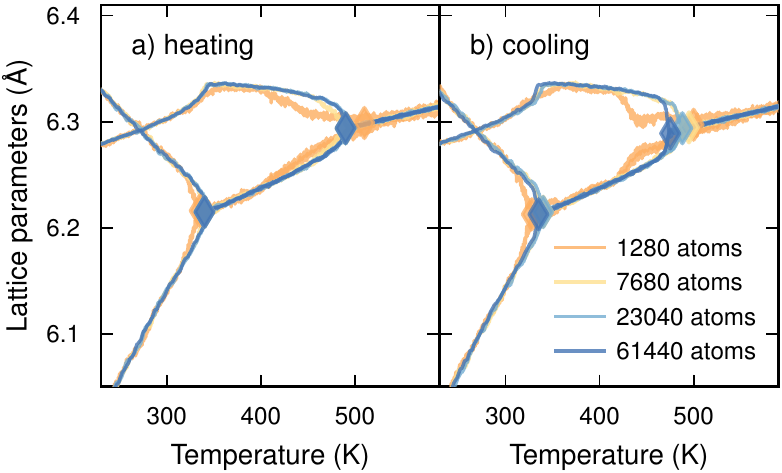}
\includegraphics{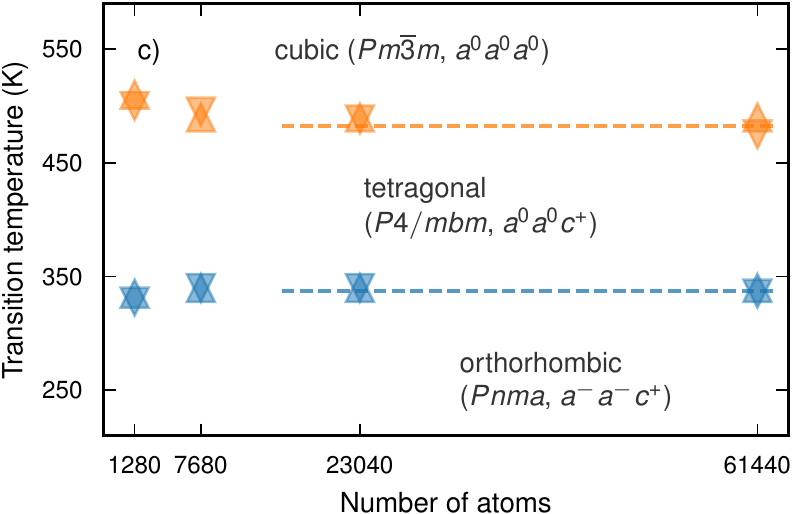}
\caption{
    \textbf{Convergence with supercell size.}
    Lattice parameters for \ce{CsPbI3} as a function of temperature from \gls{md} simulations during (a) heating and (b) cooling at a rate of $R=\SI{6}{\kelvin\per\nano\second}$ for different supercells.
    The transition temperatures extracted from these data are indicated by diamonds.
    A Hamming window of \SI{0.6}{\kelvin} was applied.
    (c) Transition temperatures as a function of supercell size.
    All results were obtained using the full model (Sect.~\nameref{sect:model-construction}) based on the \gls{cx} exchange-correlation functional.
}
\label{fig:size-effects}
\end{figure}

The temperatures for the transitions between the perovskite phases can be readily obtained from the lattice parameters (\autoref{fig:rate-effects}a,b), revealing a strong dependence on the heating/cooling rate.
For the rates below approximately \SI{30}{\kelvin\per\nano\second}, the hysteresis between heating and cooling runs is \SI{15}{\kelvin} are less and no longer vary systematically with the rate (\autoref{fig:rate-effects}c).
By contrast, for the largest rate of \SI{600}{\kelvin\per\nano\second} considered here, which is only slightly larger than values used in some earlier \gls{mlp} studies \cite{jinnouchi2019phase, LahBok22}, one observes a hysteresis of \SI{100}{\kelvin} or more for both the lower and higher temperature transitions.
In addition, the transition itself is smeared out in temperature, which is particular apparent for the orthorhombic-to-tetragonal transition (\autoref{fig:rate-effects}a).

We observed similar trends also for the other materials and models studied in this work.
We therefore conclude that rates below approximately \SI{30}{\kelvin\per\nano\second} are recommended in order to achieve convergence of the transition temperatures for this class of materials.

\subsubsection*{Size effects}
\label{sect:size-effects}

Next we examine the impact of supercell size on the temperature dependence of the lattice parameters and the transition temperatures.
First, simulations were carried out at a heating/cooling rate of \SI{6}{\kelvin\per\nano\second} using structures comprising \num{1280} atoms (4x4x4 primitive orthorhombic unit cells), \num{7680} atoms (8x8x6), \num{23040} atoms (12x12x8) or \num{61440} atoms (16x16x12).

\begin{figure}[tb]
\centering
\includegraphics{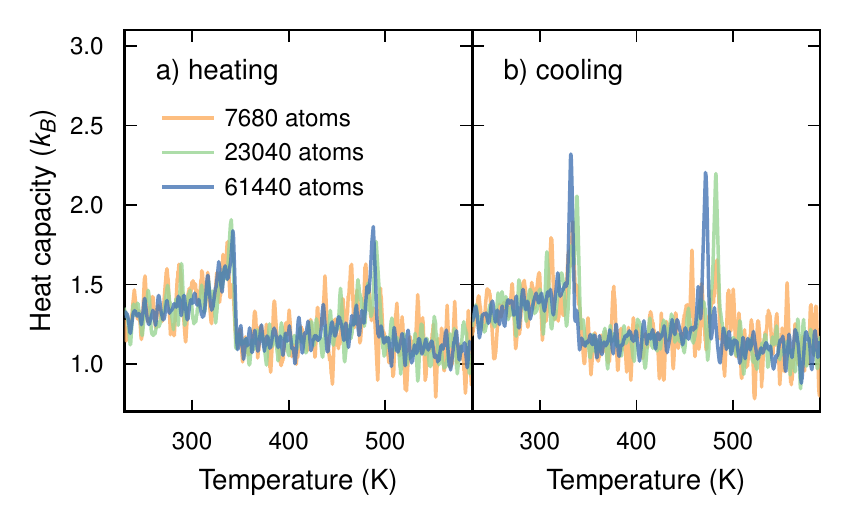}
\caption{
    \textbf{Convergence of heat capacity with supercell size.}
    Isobaric heat capacity of \ce{CsPbI3} as a function of temperature for different supercell sizes obtained through numerical differentiation as detailed in the text.
    The phase transitions are visible as peaks in the heat capacity.
    All results were obtained using the full model (Sect.~\nameref{sect:model-construction}) based on the \gls{cx} exchange-correlation functional.
}
\label{fig:heat-capacity-window-size-system-size}
\end{figure}

For the smallest supercell (\num{7680} atoms) one notices a marked deviation from the (reference) lattice parameter parameter data from the largest supercell (\num{61440} atoms) around the cubic-tetragonal phase transition (\autoref{fig:size-effects}a,b).
This deviation is, however, absent for the next larger structure of \num{7680} atoms and at this size the transition temperatures are already converged to within \SI{10}{\kelvin} of the results for the largest supercell (\autoref{fig:size-effects}c).

A key characteristic of a phase transition that is at least partly continuous is a peak or kink in the heat capacity.
The heat capacity can be readily extracted from the fluctuations of the energy in \gls{md} simulations.
For the present purpose this is, however, impractical as the transition is very sharp in temperature and the temperature range of interest is wide.
Here, we therefore compute the heat capacity instead by numerically differentiating the potential energy from heating/cooling runs, i.e., $C_p=dH/d T\approx \Delta H/\Delta T$.
This requires averaging of the data in order to obtain numerically well behaved results.
To this end, we first apply a Hamming window of \SI{0.6}{\kelvin} to the energy-vs-temperature data.
The resulting data is numerically differentiated, after which the data is smoothened again using a Hamming window of \SI{6}{\kelvin}.
The Hamming window sizes are chosen to be sufficiently large to remove noise and small enough to avoid removing features (\autoref{sfig:heat-capacity-window-size}).

For the largest system size (\num{61440} atoms) the phase transitions are clearly visible as peaks in the temperature dependence of the heat capacity (\autoref{fig:heat-capacity-window-size-system-size}).
These features become, however, less distinct with decreasing system size as fluctuations increase.

The analysis presented in this section suggests that supercells with at least about \num{10000} atoms can be expected to yield accurate lattice parameters and transition temperatures within about \SI{10}{\kelvin} of the converged results.
Extracting the temperature dependence of the heat capacity requires larger systems still.
Even for the largest systems considered here (\num{61440} atoms) the noise level is rather high, but the data still allows one to accurately extract phase transition temperatures from the heat capacity data.

\subsection*{Model uncertainty}
\label{sect:model-uncertainty}

\begin{figure}
\centering
\includegraphics{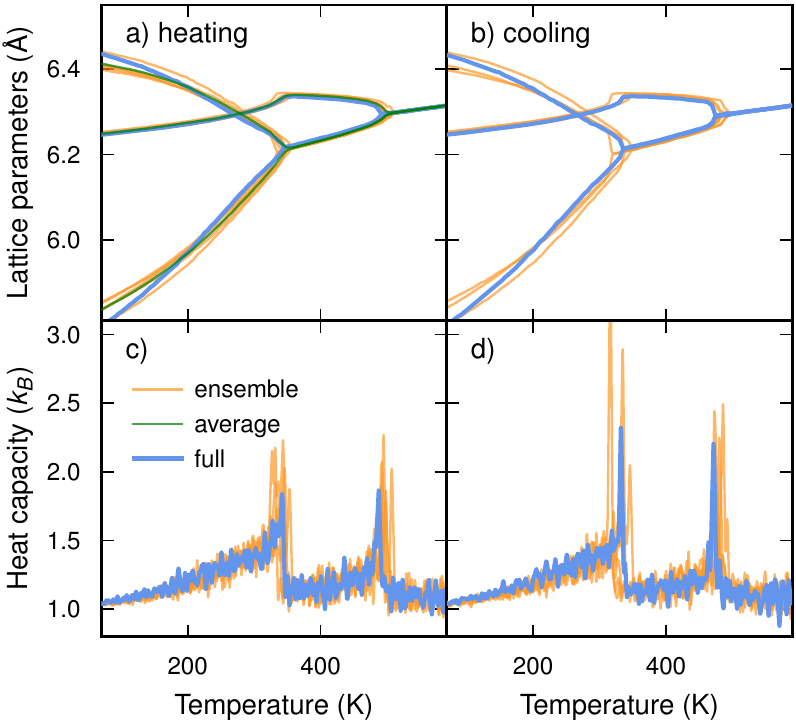}
\caption{
    \textbf{Assessing model uncertainty through ensemble of models.}
    (a,b) Lattice parameters and (c,d) heat capacity as a function of temperature during (a,c) heating and (b,d) cooling from full (blue lines) and ensemble models (orange lines).
    All results are for \ce{CsPbI3} using models based on the \gls{cx} exchange-correlation functional.
}
\label{fig:lattice-parameters-uncertainty}
\end{figure}

Having established heating/cooling rates and system sizes that yield converged results for lattice parameters and transition temperatures, we can now address the model uncertainty.
To this end, we resort to ensemble models.
The latter comprise five separate models constructed using five distinct 90-10 splits of the training data (Sect.~\nameref{sect:model-construction}).
All simulations in this section were carried out using supercells with \num{61440} atoms and a heating/cooling rate of \SI{6}{\kelvin\per\nano\second}.
Once again we use \ce{CsPbI3} and models trained using reference data generated by the \gls{cx} functional as a representative example.

The uncertainty of the model predictions can be estimated by the considering the standard deviation over the model ensemble (\autoref{fig:lattice-parameters-uncertainty}a,b).
At \SI{100}{\kelvin} this approach yields an uncertainty of up to \SI{0.02}{\angstrom} depending on direction.
This value diminishes, however, quickly with temperature to a level of \SI{0.003}{\angstrom} in the tetragonal and even less than \SI{0.002}{\angstrom} in the cubic phases (\autoref{sfig:uncertainty-lattice-parameters}).

Away from the phase transitions the heat capacity curves from the different models agree well with each other (\autoref{fig:lattice-parameters-uncertainty}).
The actual transition temperatures, corresponding to the position of the peaks in the heat capacity curves, obtained from the full model (and the model ensemble) are \SI{340}{\kelvin} (\SI{341+-9}{\kelvin}) and \SI{487}{\kelvin} (\SI{496+-7}{\kelvin}) on heating and \SI{331}{\kelvin} (\SI{323+-17}{\kelvin}) and \SI{470}{\kelvin} (\SI{478+-5}{\kelvin}) on cooling.
The agreement between the results obtained using the full model and the model ensemble support the good convergence of the models with respect to training data.
The remaining hysteresis can be attributed to the mixed first/continuous-order character, which is evident from the very small but non-zero latent heat associated with these transitions \cite{Hirotsu1974, Rodov2003, Malyshkin2020, Klarbring2019}.

\begin{figure}
\centering
\includegraphics{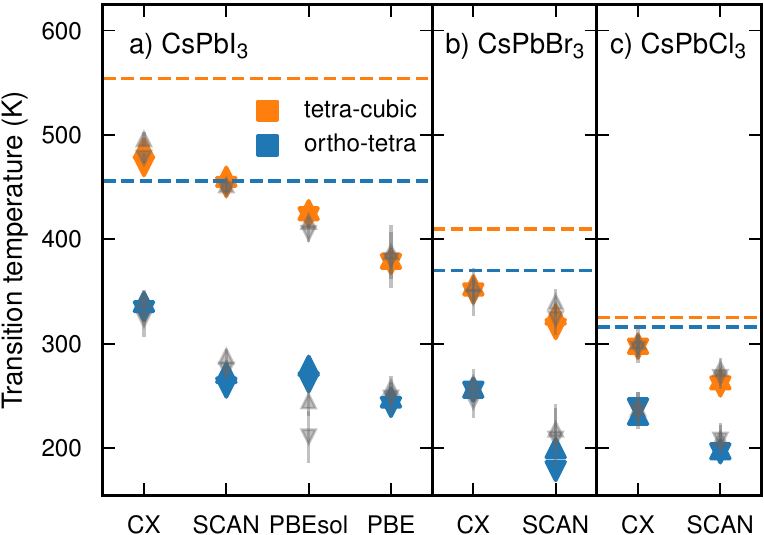}
\caption{
    \textbf{Transition temperatures from \gls{md} simulations in comparison with experiment.}
    Data were obtained using heating (upward triangles) and cooling rates (downward triangles) of \SI{6}{\nano\second} for supercells comprising \num{61440} atoms.
    Colored and gray symbols indicate results obtained using full models and model ensembles, respectively.
    In the latter case the uncertainty calculated as the standard deviation over the ensemble is indicated by vertical bars.
    Experimental transition temperatures taken from Refs.~\citenum{marronnier2018anharmonicity}, \citenum{Stoumpos2013}, and \citenum{HeStoHad21} are shown by horizontal dashed lines.
}
\label{fig:transition-temperatures-comparison-of-functionals}
\end{figure}

We can thus conservatively estimate the error in the transition temperatures due to model uncertainty to be on the order of \SI{20}{\kelvin}.
In combination with the model performance measures (Sect.~\autoref{sect:model-construction}) and the very good agreement with the \gls{dft} reference data, this provides strong evidence that the \gls{nep} models constructed are accurate representations of the \gls{dft} potential energy landscape in the regions of configuration space included here.
They can thus be used to analyze the performance of different \gls{xc} functionals with respect to the finite temperature behavior of \ce{CsPbI3} and the other materials considered here (Sect.~\nameref{sect:comparison-of-functionals}).

\subsection*{Impact of \texorpdfstring{\gls{xc}}{XC} functional and extension to other halides}
\label{sect:xc-functional}
\label{sect:comparison-of-functionals}

\begin{figure}
\centering
\includegraphics{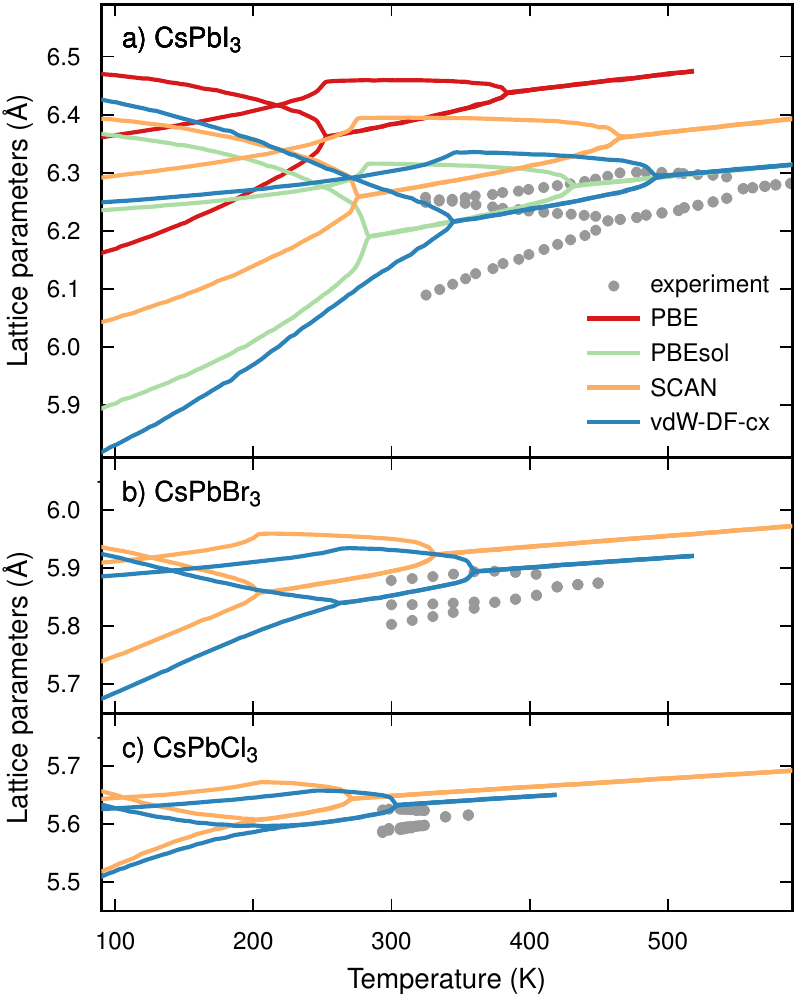}
\caption{
    \textbf{Lattice parameters as a function of temperature from simulation in comparison with experiment.}
    Data were obtained using ``full'' \gls{nep} models trained using different \gls{xc} functionals in comparison.
    Experimental data from Refs.~\citenum{marronnier2018anharmonicity}, \citenum{Stoumpos2013}, and \citenum{HeStoHad21}.
}
\label{fig:lattice-parameter-comparison}
\end{figure}

We can now apply the framework established above to predict transition temperatures for \ce{CsPbI3} using different \gls{xc} functionals for generating reference data, and compare the results to experimental data.
To this end, we use the same computational settings in terms of system size and heating/cooling rate as in the analysis of the model uncertainty.

The results show that all \gls{xc} functionals considered here systematically underestimate the transition temperatures for both the orthorhombic-tetragonal and tetragonal-cubic transitions by as much as \SI{200}{\kelvin} (\autoref{fig:transition-temperatures-comparison-of-functionals}).
The agreement improves in the sequence PBE, PBEsol, \gls{scan}, and \gls{cx}, which mirrors the trends observed previously for transition metals \cite{GhaErhHyl17}.

Conversely for the lattice parameters one observes a systematic overestimation relative to experiment (\autoref{fig:lattice-parameter-comparison}) with a similar trend.
The closest agreement is obtained using the \gls{cx} functional, for which the lattice parameters of the cubic phase are still overestimated by \num{0.02} to \SI{0.04}{\angstrom} in the high-temperature limit.

Finally, we extend our investigation to \ce{CsPbBr3} and \ce{CsPbCl3} including the \gls{scan} and \gls{cx} functionals.
As for \ce{CsPbI3} the transition temperatures (\autoref{fig:transition-temperatures-comparison-of-functionals}b,c) and lattice parameters (\autoref{fig:lattice-parameter-comparison}b,c) are systematically underestimated and overestimated, respectively.
The thermal expansion is, however, captured well by all functionals.

We conclude that while the \gls{xc} functionals considered here yield the correct sequence of phases, none of them are quantitatively in satisfactory agreement with experiment, systematically underestimating the transition temperatures (\autoref{fig:transition-temperatures-comparison-of-functionals}) and overestimating the lattice parameters (\autoref{fig:lattice-parameter-comparison}).

\subsection*{Transition to \texorpdfstring{$\delta$}{δ}-phase}
\label{sect:delta-phase}

It is experimentally well established that the perovskite phases of \ce{CsPbI3} are only metastable at low temperatures as the actual ground state structure of the material is the so-called $\delta$-phase \cite{SutFilHag18}.
We therefore also computed the transition temperature between the cubic perovskite and the $\delta$-phase by free energy integration (see \autoref{sfig:free-energies-delta} for the free energy curves).

The transition temperature obtained by the \gls{cx} model of \SI{635}{\kelvin} with an estimated uncertainty of \num{20} to \SI{40}{\kelvin} is in rather good agreement with experimental values of around \SI{600}{\kelvin} (\autoref{tab:delta-cubic-transition}).
The \gls{scan} and PBEsol models yield lower values that are considerably smaller than the experimental data.

According to \gls{dft} (and \gls{nep}) calculations the $\delta$-phase is the most stable structure for \ce{CsPbBr3}.
The energy difference (\autoref{stab:model-predictions-CsPbBr3}) is, however, much smaller than for \ce{CsPbI3} (\autoref{stab:model-predictions-CsPbI3-1}), leading to lower transition temperatures that are predicted to be \SI{310}{\kelvin} and \SI{151}{\kelvin} according to the \gls{cx} and \gls{scan} models, respectively.
We are unaware of experimental measurements of the transition temperature, which given the present predictions might actually be close to or below room temperature and thus difficult to observe.

For \ce{CsPbCl3} the \gls{dft} calculations yield energy differences between perovskite and $\delta$-phases close to zero (\autoref{stab:model-predictions-CsPbCl3}), suggesting that the $\delta$-phase is actually not the most stable phase under any conditions or only at extremely low temperatures.

\begin{table}
\centering
\caption{
    Temperatures for the transition between the $\delta$-phase and the cubic perovskite phase.
    Experimental data from Refs.~\citenum{DasHawDil17, MarRomBoy18, KeWanJia21}.
}
\label{tab:delta-cubic-transition}
\begin{tabular}{l*{5}r}
\toprule
& \multicolumn{1}{c}{\gls{cx}}
& \multicolumn{1}{c}{\gls{scan}}
& \multicolumn{1}{c}{PBEsol}
& \multicolumn{1}{c}{PBE}
& \multicolumn{1}{c}{Experiment}
\\
\midrule
\ce{CsPbI3}  &   635 & 432 & 380 & 310 & $\sim$600 \\
\ce{CsPbBr3} &   310 & 151 &     &     &           \\
\bottomrule
\end{tabular}
\end{table}

\section*{Discussion}
\label{sect:discussion}

We have systematically analyzed four key sources of error in atomic scale simulations of phase transitions of inorganic halide perovskites, related respectively to (1) sampling time, (2) system size, (3) model uncertainty, and (4) the underlying \gls{xc} functional.
Based on these results, it is recommended to use heating/cooling rates of at most approximately \SI{30}{\kelvin\per\nano\second} but preferably even lower and system sizes comprising at least a couple ten thousand atoms, corresponding to a few thousand primitive unit cells.
We expect that these guidelines are not limited to inorganic halide perovskites but should also be heeded when modeling the dynamics of other perovskites and related systems.

The model uncertainty was assessed using ensembles of models, from which uncertainty estimates for, e.g., transition temperatures and lattice parameters can be estimated.
The results show that by careful model construction the model uncertainty can be reduced to a level that allows one to quantitatively discriminate the performance of different \gls{xc} functionals.

Using this approach, we found that the semi-local \gls{xc} functionals considered here systematically underestimate the transition temperatures and overestimate the lattice parameters at finite temperature compared to experiment.
The best overall agreement is obtained for the \gls{cx} functional, which also outperforms the \gls{scan} functional.
In pioneering work on the use of \glspl{mlp} for probing the dynamics of halide perovskite, the latter functional had been suggested as achieving a good match with experiment \cite{PauSunPer17, ZhaSunPer17, jinnouchi2019phase, LahBok22}.
Our analysis suggests that the good agreement was likely fortuitous and likely the result of using high rates and small system sizes (see \autoref{sfig:lattice-parameters-comparison-GAP} for a more detailed comparison).

This raises the question how, for example, hybrid functionals or the random phase approximation would perform with regard to the finite temperature properties and dynamics of these materials \cite{BokLahRam17, VerRanFra22, BraGoeVan22}.
In the present work we included a relatively large set of structures and supercells that would pose a considerable computational challenge for either one of these methods.
As noted above (Sect.~\nameref{sect:model-construction}), one can, however, expect that the number of training structures can be considerably reduced without a notable loss in model accuracy by using active learning.
This strategy could be combined with principal component analysis to identify regions with very dense sampling \cite{FanWanYin22} or entropy maximization \cite{KarPer20} to reduce the training set size even further, eventually allowing one to build \gls{nep} or other \gls{mlp} models that can represent such more accurate electronic structure methods.

Finally, we note that the accuracy of the models presented here in combination with the very high computational efficiency provide by the implementation on \glspl{gpu}, now enables one to sample the dynamics of these and related materials with unprecedented time resolution and accuracy.

\section*{Methods}

\subsection*{Machine learned potentials}

\subsubsection*{Neuroevolution potentials}

Here, we use the third generation of the \gls{nep} scheme (NEP3) \cite{FanWanYin22} to build \glspl{mlp} for \ce{CsPbX3} with X=Cl, Br, and I.
The \gls{nep} format employs a simple multi-layer perceptron neural network architecture with a single hidden layer \cite{FanZenZha21}.
In NEP3 the radial part of the atomic environment descriptor is constructed from linear combinations of Chebyshev basis functions while the three-body angular part is similarly build from Legendre polynomials.
Four and five-body terms of the atomic cluster expansion form \cite{Dra19} can be included as well but here we limit ourselves to two and three-body terms.

For the present purpose it is crucial that the \gls{nep} scheme is not only accurate but has been implemented on \glspl{gpu} in the \textsc{gpumd} package \cite{FanWanYin22}.
For the models described in the following this allows us to achieve a speed of \SI{2e7}{\atom\step\per\second} on an NVidia A100 card, i.e., we can simulate a system of \num{60000} atoms for about \SI{150}{\nano\second} per day using a time step of \SI{5}{\femto\second}.

\subsubsection*{Computational parameters}

In this study we used the same hyperparameters for all models, which were chosen based on experience and pre-trials \cite{FanWanYin22}.
The cutoffs for two and three-body interactions are \SI{8}{\angstrom} and \SI{4}{\angstrom}, respectively.
There are 8 radial and 6 angular descriptor components, 8 basis functions for building both the radial and angular descriptor functions, and the angular components are expanded up to fourth order.
The hidden layer contains 50 neurons.

The weights for energies, forces, and virials in the loss function were set to 1, 5, and 0.2 in \textsc{gpumd} units, respectively, while the weights for the $\ell_1$ and $\ell_2$ regularization terms were set to 0.1.
The neuroevolution strategy \cite{WieSchGla14} used for optimizing the parameters used a population size of 50 and was run for \num{200000} generations.

\subsubsection*{Model construction}
\label{sect:model-construction}

To construct \gls{nep} models we employed a boot-strapping strategy.
First we identified potentially relevant phases.
This included the cubic perovskite structure ($Pm\bar{3}m$, Glazer notation $a^0a^0a^0$), two tetragonal structures ($I4/mcm$ $\rightarrow$ $a^0a^0c^-$, $P4/mbm$ $\rightarrow$ $a^0a^0c^+$), representing out-of-phase and in-phase tilts relative to the $c$-axis, respectively, one orthorhombic structure ($Pnma$ $\rightarrow$ $a^-a^-c^+$) as well as the so-called delta-phase ($Pnma$), which is experimentally known to be the most stable structure at least for \ce{CsPbI3} and \ce{CsPbBr3}.

We then calculated energy-volume curves for these five prototype structures using \gls{dft} calculations (Sect.~\nameref{sect:dft-calculations}) allowing both the ionic coordinates and the cell shape to relax under the constraint of constant volume until the maximum force on any atom fell below \SI{30}{\milli\electronvolt\per\angstrom}.

Subsequently we generated supercells for each prototype with random atomic displacements using the Monte Carlo rattle procedure from the \textsc{hiphive} package \cite{EriFraErh19} with a standard deviation of \SI{0.04}{\angstrom}.
The supercell size was chosen to be between 120 and 160 atoms and the volume was varied between 85\% and 110\% of the respective equilibrium volume with five structures per volume and prototype.

Using these data we generated a first iteration of \gls{nep} models using the \textsc{gpumd} package for the optimization \cite{FanWanYin22} and the \textsc{calorine} package for data preparation and analysis \cite{calorine}.
One model was generated using the full data set (``full model'') and five additional models (``model ensemble'') were generated by using five different 90-10 splits of the available data.
Using the full model we generated new structures for each prototype by running short \gls{md} simulations at pressures between \num{-1} and \SI{10}{\giga\pascal} using a temperature ramp from \num{20} to \SI{620}{\kelvin} over \SI{3}{\nano\second}.
From each trajectory we selected 12 configurations.
For each of these configurations we then computed the standard deviation of the energy and forces using the model ensemble.
The standard deviation over the ensemble predictions provided a measure for the uncertainty of the current model generation for the respective conditions (temperature, pressure, structure).
We then computed energy and forces for the new structures using \gls{dft} calculations, added these to the training set and repeated the procedure.
Typically after four generations we found that the uncertainty in the energy and forces was comparable are smaller than the respective training error indicating convergence of the model construction.

We note that in principle one could have adapted an active learning strategy based on the model ensemble and only included configurations with high uncertainty as additional reference structures.
Here, we decided to include rather more data in the training set but we expect that the number of structures can be reduced considerably without a notable decrease in model performance.

\begin{table}
\centering
\caption{
    \Gls{rmse} scores for the final \gls{nep} models obtained by training against the full data set.
    Additional performance measures including \gls{rmse} and Pearson correlation coefficients for model ensembles can be found in \autoref{stab:model-performance}.
}
\label{tab:models}
\begin{tabularx}{0.95\columnwidth}{lXddd}
\toprule
&
& \multicolumn{1}{c}{Energy}
& \multicolumn{1}{c}{Force}
& \multicolumn{1}{c}{Virial}
\\
&
& \multicolumn{1}{c}{meV/atom} %(\unit{\milli\electronvolt\per\atom})}
& \multicolumn{1}{c}{meV/Å} %(\unit{\milli\electronvolt\per\angstrom})}
& \multicolumn{1}{c}{meV/atom} %(\unit{\milli\electronvolt\per\atom})}
\\
\midrule
\multicolumn{5}{l}{\ce{CsPbCl3}} \\
& vdW-DF-cx  &   1.2 &  46.4 &  13.0 \\
& SCAN       &   1.1 &  47.8 &  14.0 \\[6pt]
\multicolumn{5}{l}{\ce{CsPbBr3}} \\
& vdW-DF-cx  &   1.1 &  45.1 &  11.1 \\
& SCAN       &   1.2 &  47.6 &  15.1 \\[6pt]
\multicolumn{5}{l}{\ce{CsPbI3}} \\
& vdW-DF-cx  &   1.8 &  47.5 &  14.5 \\
& SCAN       &   2.1 &  51.4 &  16.0 \\
& PBEsol     &   1.9 &  50.6 &  15.2 \\
& PBE        &   1.0 &  43.1 &  12.6 \\
\bottomrule
\end{tabularx}
\end{table}

The final models yield \gls{rmse} scores for training and validation sets of about \SI{2}{\milli\electronvolt\per\atom}, \SI{50}{\milli\electronvolt\per\angstrom}, and \SI{15}{\milli\electronvolt\per\atom} or better for energies, forces, and virials, respectively (\autoref{tab:models} and \autoref{stab:model-performance}).
Importantly the models closely reproduce the energy differences and energy-volume curves of all the structures of interest in the present study (\autoref{sfig:loss-curves-CsPbI3-cx}; \autoref{sfig:parity-plots-CsPbI3-cx}; \autoref{stab:model-predictions-CsPbCl3} to \autoref{stab:model-predictions-CsPbI3-2}).
The final models were subsequently used in large scale \gls{md} simulations to predict, for example, transition temperatures or lattice parameters (Sect.~\nameref{sect:md-simulations}).

\subsection*{\texorpdfstring{\Gls{md}}{MD} simulations}
\label{sect:md-simulations}

All \gls{md} simulations were carried out using the \textsc{gpumd} code.
Temperature and pressure were controlled using stochastic velocity \cite{BusDonPar07} and cell rescaling \cite{BerBus20} and the time step was \SI{5}{\pico\second}, where all simulations were run at zero pressure.

For studying the convergence with size (Sect.~\nameref{sect:size-effects}), we considered system sizes between \num{1280} and \num{61440} atoms, equivalent to 4x4x4 to 16x16x12 primitive orthorhombic perovskite (20-atom) unit cells.
To analyze the impact of heating and cooling rates (Sect.~\nameref{sect:rate-effects}) the temperature was linearly varied between \SI{20}{\kelvin} and \SI{620}{\kelvin} over \SI{1}{\nano\second} to \SI{100}{\nano\second}.

The production runs used to quantify model uncertainty (Sect.~\nameref{sect:model-uncertainty}) and the impact of the \gls{xc} functional (Sect.~\nameref{sect:xc-functional}) were carried out using supercells comprising 16x16x12 primitive orthorhombic unit cells (\num{61440} atoms).
The total simulation time was set to \SI{100}{\nano\second} and the temperature was varied over a range of \num{400} to \SI{600}{\kelvin} corresponding to a heating/cooling rate of \num{4} to \SI{6}{\kelvin\per\nano\second}.

\subsection*{Free energy calculations}

For \ce{CsPbI3} and possibly \ce{CsPbBr3} the perovskite phases are only metastable at lower temperatures.
Provided sufficient kinetic activation, below a certain temperature the perovskite structure transforms into the so-called $\delta$-phase via a first order transition.
To determine the transition temperature from the \gls{nep} models we calculated the free energies of the $\delta$ and cubic perovskite phases through thermodynamic integration using the classical method by Frenkel and Ladd \cite{FreLad84, FreSmi01, FreAstde16}, as implemented in \textsc{ase} \cite{LarMorBlo17}. 
In these calculations, we used an Einstein solid as reference system, for which the free energy can be computed analytically, and used supercells containing about \num{1500} atoms for each phase.
For each temperature the integration  was carried out over \SI{50}{ps} and the results were averaged over ten independent runs.

\subsection*{Reference calculations}
\label{sect:dft-calculations}

\Gls{dft} calculations were performed using the projector augmented-wave method \cite{Blo94} as implemented in the Vienna ab-initio simulation package \cite{KreHaf93, KreFur96}.
The exchange-correlation contribution was represented using the \gls{cx} method \cite{BerHyl2014}, the \gls{scan} density functional \cite{SunRuzPer15}, the PBEsol functional \cite{PerRuz2008}, and the PBE functional \cite{PerBurErn}.
The Brillouin zone was sampled with a $\Gamma$-centered grid with a $\vec{k}$-point density of \SI{0.18}{\per\angstrom} and Gaussian smearing with a width of \SI{0.1}{\electronvolt}.
For the calculation of the forces a finer support grid was employed to improve their numerical accuracy.

\begin{acknowledgments}
This work was funded by the Swedish Research Council (grant numbers 2018-06482, 2019-03993, 2020-04935, 2021-05072) and the Chalmers Initiative for Advancement of Neutron and Synchrotron Techniques.
The computations were enabled by resources provided by the Swedish National Infrastructure for Computing (SNIC) at NSC, C3SE, and PDC partially funded by the Swedish Research Council (grant number 2018-05973).
\end{acknowledgments}

\section*{Data Availability}

The \gls{dft} data and \gls{nep} models generated in this study are openly available via Zenodo at \url{https://doi.org/10.5281/zenodo.7454224}.

\section*{Competing interests}

The authors declare no competing interests.

\section*{Author Contributions}

All authors contributed equally.


\begin{thebibliography}{63}%
\makeatletter
\providecommand \@ifxundefined [1]{%
 \@ifx{#1\undefined}
}%
\providecommand \@ifnum [1]{%
 \ifnum #1\expandafter \@firstoftwo
 \else \expandafter \@secondoftwo
 \fi
}%
\providecommand \@ifx [1]{%
 \ifx #1\expandafter \@firstoftwo
 \else \expandafter \@secondoftwo
 \fi
}%
\providecommand \natexlab [1]{#1}%
\providecommand \enquote  [1]{``#1''}%
\providecommand \bibnamefont  [1]{#1}%
\providecommand \bibfnamefont [1]{#1}%
\providecommand \citenamefont [1]{#1}%
\providecommand \href@noop [0]{\@secondoftwo}%
\providecommand \href [0]{\begingroup \@sanitize@url \@href}%
\providecommand \@href[1]{\@@startlink{#1}\@@href}%
\providecommand \@@href[1]{\endgroup#1\@@endlink}%
\providecommand \@sanitize@url [0]{\catcode `\\12\catcode `\$12\catcode
  `\&12\catcode `\#12\catcode `\^12\catcode `\_12\catcode `\%12\relax}%
\providecommand \@@startlink[1]{}%
\providecommand \@@endlink[0]{}%
\providecommand \url  [0]{\begingroup\@sanitize@url \@url }%
\providecommand \@url [1]{\endgroup\@href {#1}{\urlprefix }}%
\providecommand \urlprefix  [0]{URL }%
\providecommand \Eprint [0]{\href }%
\providecommand \doibase [0]{https://doi.org/}%
\providecommand \selectlanguage [0]{\@gobble}%
\providecommand \bibinfo  [0]{\@secondoftwo}%
\providecommand \bibfield  [0]{\@secondoftwo}%
\providecommand \translation [1]{[#1]}%
\providecommand \BibitemOpen [0]{}%
\providecommand \bibitemStop [0]{}%
\providecommand \bibitemNoStop [0]{.\EOS\space}%
\providecommand \EOS [0]{\spacefactor3000\relax}%
\providecommand \BibitemShut  [1]{\csname bibitem#1\endcsname}%
\let\auto@bib@innerbib\@empty
%</preamble>
\bibitem [{\citenamefont {Kojima}\ \emph {et~al.}(2009)\citenamefont {Kojima},
  \citenamefont {Teshima}, \citenamefont {Shirai},\ and\ \citenamefont
  {Miyasaka}}]{kojima2009organometal}%
  \BibitemOpen
  \bibfield  {author} {\bibinfo {author} {\bibfnamefont {A.}~\bibnamefont
  {Kojima}}, \bibinfo {author} {\bibfnamefont {K.}~\bibnamefont {Teshima}},
  \bibinfo {author} {\bibfnamefont {Y.}~\bibnamefont {Shirai}},\ and\ \bibinfo
  {author} {\bibfnamefont {T.}~\bibnamefont {Miyasaka}},\ }\href
  {https://doi.org/10.1021/ja809598r} {\bibfield  {journal} {\bibinfo
  {journal} {Journal of the American Chemical Society}\ }\textbf {\bibinfo
  {volume} {131}},\ \bibinfo {pages} {6050} (\bibinfo {year}
  {2009})}\BibitemShut {NoStop}%
\bibitem [{\citenamefont {Kim}\ \emph {et~al.}(2012)\citenamefont {Kim},
  \citenamefont {Lee}, \citenamefont {Im}, \citenamefont {Lee}, \citenamefont
  {Moehl}, \citenamefont {Marchioro}, \citenamefont {Moon}, \citenamefont
  {Humphry-Baker}, \citenamefont {Yum}, \citenamefont {Moser} \emph
  {et~al.}}]{kim2012lead}%
  \BibitemOpen
  \bibfield  {author} {\bibinfo {author} {\bibfnamefont {H.-S.}\ \bibnamefont
  {Kim}}, \bibinfo {author} {\bibfnamefont {C.-R.}\ \bibnamefont {Lee}},
  \bibinfo {author} {\bibfnamefont {J.-H.}\ \bibnamefont {Im}}, \bibinfo
  {author} {\bibfnamefont {K.-B.}\ \bibnamefont {Lee}}, \bibinfo {author}
  {\bibfnamefont {T.}~\bibnamefont {Moehl}}, \bibinfo {author} {\bibfnamefont
  {A.}~\bibnamefont {Marchioro}}, \bibinfo {author} {\bibfnamefont {S.-J.}\
  \bibnamefont {Moon}}, \bibinfo {author} {\bibfnamefont {R.}~\bibnamefont
  {Humphry-Baker}}, \bibinfo {author} {\bibfnamefont {J.-H.}\ \bibnamefont
  {Yum}}, \bibinfo {author} {\bibfnamefont {J.~E.}\ \bibnamefont {Moser}},
  \emph {et~al.},\ }\href {https://doi.org/10.1038/srep00591} {\bibfield
  {journal} {\bibinfo  {journal} {Scientific Reports}\ }\textbf {\bibinfo
  {volume} {2}},\ \bibinfo {pages} {1} (\bibinfo {year} {2012})}\BibitemShut
  {NoStop}%
\bibitem [{\citenamefont {Hodes}(2013)}]{hodes2013perovskite}%
  \BibitemOpen
  \bibfield  {author} {\bibinfo {author} {\bibfnamefont {G.}~\bibnamefont
  {Hodes}},\ }\href {https://doi.org/10.1126/science.124547} {\bibfield
  {journal} {\bibinfo  {journal} {Science}\ }\textbf {\bibinfo {volume}
  {342}},\ \bibinfo {pages} {317} (\bibinfo {year} {2013})}\BibitemShut
  {NoStop}%
\bibitem [{\citenamefont {Van~Le}\ \emph {et~al.}(2018)\citenamefont {Van~Le},
  \citenamefont {Jang},\ and\ \citenamefont {Kim}}]{van2018recent}%
  \BibitemOpen
  \bibfield  {author} {\bibinfo {author} {\bibfnamefont {Q.}~\bibnamefont
  {Van~Le}}, \bibinfo {author} {\bibfnamefont {H.~W.}\ \bibnamefont {Jang}},\
  and\ \bibinfo {author} {\bibfnamefont {S.~Y.}\ \bibnamefont {Kim}},\ }\href
  {https://doi.org/10.1002/smtd.201700419} {\bibfield  {journal} {\bibinfo
  {journal} {Small Methods}\ }\textbf {\bibinfo {volume} {2}},\ \bibinfo
  {pages} {1700419} (\bibinfo {year} {2018})}\BibitemShut {NoStop}%
\bibitem [{\citenamefont {Tyson}\ \emph {et~al.}(2017)\citenamefont {Tyson},
  \citenamefont {Gao}, \citenamefont {Chen}, \citenamefont {Ghose},\ and\
  \citenamefont {Yan}}]{tyson2017large}%
  \BibitemOpen
  \bibfield  {author} {\bibinfo {author} {\bibfnamefont {T.}~\bibnamefont
  {Tyson}}, \bibinfo {author} {\bibfnamefont {W.}~\bibnamefont {Gao}}, \bibinfo
  {author} {\bibfnamefont {Y.-S.}\ \bibnamefont {Chen}}, \bibinfo {author}
  {\bibfnamefont {S.}~\bibnamefont {Ghose}},\ and\ \bibinfo {author}
  {\bibfnamefont {Y.}~\bibnamefont {Yan}},\ }\href
  {https://doi.org/10.1038/s41598-017-09220-2} {\bibfield  {journal} {\bibinfo
  {journal} {Scientific Reports}\ }\textbf {\bibinfo {volume} {7}},\ \bibinfo
  {pages} {1} (\bibinfo {year} {2017})}\BibitemShut {NoStop}%
\bibitem [{\citenamefont {Liu}\ \emph {et~al.}(2022)\citenamefont {Liu},
  \citenamefont {Du}, \citenamefont {Phillips}, \citenamefont {Wyatt},
  \citenamefont {Keen},\ and\ \citenamefont {Dove}}]{liu2022neutron}%
  \BibitemOpen
  \bibfield  {author} {\bibinfo {author} {\bibfnamefont {J.}~\bibnamefont
  {Liu}}, \bibinfo {author} {\bibfnamefont {J.}~\bibnamefont {Du}}, \bibinfo
  {author} {\bibfnamefont {A.~E.}\ \bibnamefont {Phillips}}, \bibinfo {author}
  {\bibfnamefont {P.~B.}\ \bibnamefont {Wyatt}}, \bibinfo {author}
  {\bibfnamefont {D.~A.}\ \bibnamefont {Keen}},\ and\ \bibinfo {author}
  {\bibfnamefont {M.~T.}\ \bibnamefont {Dove}},\ }\href
  {https://doi.org/0.1088/1361-648X/ac4aa9} {\bibfield  {journal} {\bibinfo
  {journal} {Journal of Physics: Condensed Matter}\ }\textbf {\bibinfo {volume}
  {34}},\ \bibinfo {pages} {145401} (\bibinfo {year} {2022})}\BibitemShut
  {NoStop}%
\bibitem [{\citenamefont {Yaffe}\ \emph {et~al.}(2017)\citenamefont {Yaffe},
  \citenamefont {Guo}, \citenamefont {Tan}, \citenamefont {Egger},
  \citenamefont {Hull}, \citenamefont {Stoumpos}, \citenamefont {Zheng},
  \citenamefont {Heinz}, \citenamefont {Kronik}, \citenamefont {Kanatzidis}
  \emph {et~al.}}]{yaffe2017local}%
  \BibitemOpen
  \bibfield  {author} {\bibinfo {author} {\bibfnamefont {O.}~\bibnamefont
  {Yaffe}}, \bibinfo {author} {\bibfnamefont {Y.}~\bibnamefont {Guo}}, \bibinfo
  {author} {\bibfnamefont {L.~Z.}\ \bibnamefont {Tan}}, \bibinfo {author}
  {\bibfnamefont {D.~A.}\ \bibnamefont {Egger}}, \bibinfo {author}
  {\bibfnamefont {T.}~\bibnamefont {Hull}}, \bibinfo {author} {\bibfnamefont
  {C.~C.}\ \bibnamefont {Stoumpos}}, \bibinfo {author} {\bibfnamefont
  {F.}~\bibnamefont {Zheng}}, \bibinfo {author} {\bibfnamefont {T.~F.}\
  \bibnamefont {Heinz}}, \bibinfo {author} {\bibfnamefont {L.}~\bibnamefont
  {Kronik}}, \bibinfo {author} {\bibfnamefont {M.~G.}\ \bibnamefont
  {Kanatzidis}}, \emph {et~al.},\ }\href
  {https://doi.org/10.1103/PhysRevLett.118.136001} {\bibfield  {journal}
  {\bibinfo  {journal} {Physical Review Letters}\ }\textbf {\bibinfo {volume}
  {118}},\ \bibinfo {pages} {136001} (\bibinfo {year} {2017})}\BibitemShut
  {NoStop}%
\bibitem [{\citenamefont {Marronnier}\ \emph {et~al.}(2017)\citenamefont
  {Marronnier}, \citenamefont {Lee}, \citenamefont {Geffroy}, \citenamefont
  {Even}, \citenamefont {Bonnassieux},\ and\ \citenamefont
  {Roma}}]{marronnier2017structural}%
  \BibitemOpen
  \bibfield  {author} {\bibinfo {author} {\bibfnamefont {A.}~\bibnamefont
  {Marronnier}}, \bibinfo {author} {\bibfnamefont {H.}~\bibnamefont {Lee}},
  \bibinfo {author} {\bibfnamefont {B.}~\bibnamefont {Geffroy}}, \bibinfo
  {author} {\bibfnamefont {J.}~\bibnamefont {Even}}, \bibinfo {author}
  {\bibfnamefont {Y.}~\bibnamefont {Bonnassieux}},\ and\ \bibinfo {author}
  {\bibfnamefont {G.}~\bibnamefont {Roma}},\ }\href
  {https://doi.org/10.1021/acs.jpclett.7b00807} {\bibfield  {journal} {\bibinfo
   {journal} {The Journal of Physical Chemistry Letters}\ }\textbf {\bibinfo
  {volume} {8}},\ \bibinfo {pages} {2659} (\bibinfo {year} {2017})}\BibitemShut
  {NoStop}%
\bibitem [{\citenamefont {Marronnier}\ \emph
  {et~al.}(2018{\natexlab{a}})\citenamefont {Marronnier}, \citenamefont {Roma},
  \citenamefont {Boyer-Richard}, \citenamefont {Pedesseau}, \citenamefont
  {Jancu}, \citenamefont {Bonnassieux}, \citenamefont {Katan}, \citenamefont
  {Stoumpos}, \citenamefont {Kanatzidis},\ and\ \citenamefont
  {Even}}]{marronnier2018anharmonicity}%
  \BibitemOpen
  \bibfield  {author} {\bibinfo {author} {\bibfnamefont {A.}~\bibnamefont
  {Marronnier}}, \bibinfo {author} {\bibfnamefont {G.}~\bibnamefont {Roma}},
  \bibinfo {author} {\bibfnamefont {S.}~\bibnamefont {Boyer-Richard}}, \bibinfo
  {author} {\bibfnamefont {L.}~\bibnamefont {Pedesseau}}, \bibinfo {author}
  {\bibfnamefont {J.-M.}\ \bibnamefont {Jancu}}, \bibinfo {author}
  {\bibfnamefont {Y.}~\bibnamefont {Bonnassieux}}, \bibinfo {author}
  {\bibfnamefont {C.}~\bibnamefont {Katan}}, \bibinfo {author} {\bibfnamefont
  {C.~C.}\ \bibnamefont {Stoumpos}}, \bibinfo {author} {\bibfnamefont {M.~G.}\
  \bibnamefont {Kanatzidis}},\ and\ \bibinfo {author} {\bibfnamefont
  {J.}~\bibnamefont {Even}},\ }\href {https://doi.org/10.1021/acsnano.8b00267}
  {\bibfield  {journal} {\bibinfo  {journal} {ACS Nano}\ }\textbf {\bibinfo
  {volume} {12}},\ \bibinfo {pages} {3477} (\bibinfo {year}
  {2018}{\natexlab{a}})}\BibitemShut {NoStop}%
\bibitem [{\citenamefont {Bechtel}\ \emph {et~al.}(2019)\citenamefont
  {Bechtel}, \citenamefont {Thomas},\ and\ \citenamefont {Van~der
  Ven}}]{bechtel2019finite}%
  \BibitemOpen
  \bibfield  {author} {\bibinfo {author} {\bibfnamefont {J.~S.}\ \bibnamefont
  {Bechtel}}, \bibinfo {author} {\bibfnamefont {J.~C.}\ \bibnamefont
  {Thomas}},\ and\ \bibinfo {author} {\bibfnamefont {A.}~\bibnamefont {Van~der
  Ven}},\ }\href {https://doi.org/10.1103/PhysRevMaterials.3.113605} {\bibfield
   {journal} {\bibinfo  {journal} {Physical Review Materials}\ }\textbf
  {\bibinfo {volume} {3}},\ \bibinfo {pages} {113605} (\bibinfo {year}
  {2019})}\BibitemShut {NoStop}%
\bibitem [{\citenamefont {Carignano}\ \emph {et~al.}(2015)\citenamefont
  {Carignano}, \citenamefont {Kachmar},\ and\ \citenamefont
  {Hutter}}]{carignano2015thermal}%
  \BibitemOpen
  \bibfield  {author} {\bibinfo {author} {\bibfnamefont {M.~A.}\ \bibnamefont
  {Carignano}}, \bibinfo {author} {\bibfnamefont {A.}~\bibnamefont {Kachmar}},\
  and\ \bibinfo {author} {\bibfnamefont {J.}~\bibnamefont {Hutter}},\ }\href
  {https://doi.org/10.1021/jp510568n} {\bibfield  {journal} {\bibinfo
  {journal} {The Journal of Physical Chemistry C}\ }\textbf {\bibinfo {volume}
  {119}},\ \bibinfo {pages} {8991} (\bibinfo {year} {2015})}\BibitemShut
  {NoStop}%
\bibitem [{\citenamefont {Wiktor}\ \emph {et~al.}(2017)\citenamefont {Wiktor},
  \citenamefont {Rothlisberger},\ and\ \citenamefont
  {Pasquarello}}]{wiktor2017predictive}%
  \BibitemOpen
  \bibfield  {author} {\bibinfo {author} {\bibfnamefont {J.}~\bibnamefont
  {Wiktor}}, \bibinfo {author} {\bibfnamefont {U.}~\bibnamefont
  {Rothlisberger}},\ and\ \bibinfo {author} {\bibfnamefont {A.}~\bibnamefont
  {Pasquarello}},\ }\href {https://doi.org/10.1021/acs.jpclett.7b02648}
  {\bibfield  {journal} {\bibinfo  {journal} {The Journal of Physical Chemistry
  Letters}\ }\textbf {\bibinfo {volume} {8}},\ \bibinfo {pages} {5507}
  (\bibinfo {year} {2017})}\BibitemShut {NoStop}%
\bibitem [{\citenamefont {Bokdam}\ \emph {et~al.}(2017)\citenamefont {Bokdam},
  \citenamefont {Lahnsteiner}, \citenamefont {Ramberger}, \citenamefont
  {Sch{\"a}fer},\ and\ \citenamefont {Kresse}}]{BokLahRam17}%
  \BibitemOpen
  \bibfield  {author} {\bibinfo {author} {\bibfnamefont {M.}~\bibnamefont
  {Bokdam}}, \bibinfo {author} {\bibfnamefont {J.}~\bibnamefont {Lahnsteiner}},
  \bibinfo {author} {\bibfnamefont {B.}~\bibnamefont {Ramberger}}, \bibinfo
  {author} {\bibfnamefont {T.}~\bibnamefont {Sch{\"a}fer}},\ and\ \bibinfo
  {author} {\bibfnamefont {G.}~\bibnamefont {Kresse}},\ }\href
  {https://doi.org/10.1103/PhysRevLett.119.145501} {\bibfield  {journal}
  {\bibinfo  {journal} {Physical Review Letters}\ }\textbf {\bibinfo {volume}
  {119}},\ \bibinfo {pages} {145501} (\bibinfo {year} {2017})}\BibitemShut
  {NoStop}%
\bibitem [{\citenamefont {Mladenovi{\'c}}\ and\ \citenamefont
  {Vukmirovi{\'c}}(2018)}]{mladenovic2018effects}%
  \BibitemOpen
  \bibfield  {author} {\bibinfo {author} {\bibfnamefont {M.}~\bibnamefont
  {Mladenovi{\'c}}}\ and\ \bibinfo {author} {\bibfnamefont {N.}~\bibnamefont
  {Vukmirovi{\'c}}},\ }\href {https://doi.org/10.1039/C8CP03726D} {\bibfield
  {journal} {\bibinfo  {journal} {Physical Chemistry Chemical Physics}\
  }\textbf {\bibinfo {volume} {20}},\ \bibinfo {pages} {25693} (\bibinfo {year}
  {2018})}\BibitemShut {NoStop}%
\bibitem [{\citenamefont {Zhu}\ \emph {et~al.}(2022)\citenamefont {Zhu},
  \citenamefont {Caicedo-D{\'a}vila}, \citenamefont {Gehrmann},\ and\
  \citenamefont {Egger}}]{zhu2022probing}%
  \BibitemOpen
  \bibfield  {author} {\bibinfo {author} {\bibfnamefont {X.}~\bibnamefont
  {Zhu}}, \bibinfo {author} {\bibfnamefont {S.}~\bibnamefont
  {Caicedo-D{\'a}vila}}, \bibinfo {author} {\bibfnamefont {C.}~\bibnamefont
  {Gehrmann}},\ and\ \bibinfo {author} {\bibfnamefont {D.~A.}\ \bibnamefont
  {Egger}},\ }\href {https://doi.org/10.1021/acsami.1c23099} {\bibfield
  {journal} {\bibinfo  {journal} {ACS Applied Materials \& Interfaces}\
  }\textbf {\bibinfo {volume} {14}},\ \bibinfo {pages} {22973} (\bibinfo {year}
  {2022})}\BibitemShut {NoStop}%
\bibitem [{\citenamefont {Gebhardt}\ and\ \citenamefont
  {Els{\"a}sser}(2022)}]{gebhardt2022electronic}%
  \BibitemOpen
  \bibfield  {author} {\bibinfo {author} {\bibfnamefont {J.}~\bibnamefont
  {Gebhardt}}\ and\ \bibinfo {author} {\bibfnamefont {C.}~\bibnamefont
  {Els{\"a}sser}},\ }\href {https://doi.org/10.1002/pssb.202200124} {\bibfield
  {journal} {\bibinfo  {journal} {Physica Status Solidi (b)}\ }\textbf
  {\bibinfo {volume} {259}},\ \bibinfo {pages} {2200124} (\bibinfo {year}
  {2022})}\BibitemShut {NoStop}%
\bibitem [{\citenamefont {Girdzis}\ \emph {et~al.}(2020)\citenamefont
  {Girdzis}, \citenamefont {Lin}, \citenamefont {Leppert}, \citenamefont
  {Slavney}, \citenamefont {Park}, \citenamefont {Chapman}, \citenamefont
  {Karunadasa},\ and\ \citenamefont {Mao}}]{girdzis2020revealing}%
  \BibitemOpen
  \bibfield  {author} {\bibinfo {author} {\bibfnamefont {S.~P.}\ \bibnamefont
  {Girdzis}}, \bibinfo {author} {\bibfnamefont {Y.}~\bibnamefont {Lin}},
  \bibinfo {author} {\bibfnamefont {L.}~\bibnamefont {Leppert}}, \bibinfo
  {author} {\bibfnamefont {A.~H.}\ \bibnamefont {Slavney}}, \bibinfo {author}
  {\bibfnamefont {S.}~\bibnamefont {Park}}, \bibinfo {author} {\bibfnamefont
  {K.~W.}\ \bibnamefont {Chapman}}, \bibinfo {author} {\bibfnamefont {H.~I.}\
  \bibnamefont {Karunadasa}},\ and\ \bibinfo {author} {\bibfnamefont {W.~L.}\
  \bibnamefont {Mao}},\ }\href {https://doi.org/10.1021/acs.jpclett.0c03412}
  {\bibfield  {journal} {\bibinfo  {journal} {The Journal of Physical Chemistry
  Letters}\ }\textbf {\bibinfo {volume} {12}},\ \bibinfo {pages} {532}
  (\bibinfo {year} {2020})}\BibitemShut {NoStop}%
\bibitem [{\citenamefont {Cannelli}\ \emph {et~al.}(2022)\citenamefont
  {Cannelli}, \citenamefont {Wiktor}, \citenamefont {Colonna}, \citenamefont
  {Leroy}, \citenamefont {Puppin}, \citenamefont {Bacellar}, \citenamefont
  {Sadykov}, \citenamefont {Krieg}, \citenamefont {Smolentsev}, \citenamefont
  {Kovalenko} \emph {et~al.}}]{cannelli2022atomic}%
  \BibitemOpen
  \bibfield  {author} {\bibinfo {author} {\bibfnamefont {O.}~\bibnamefont
  {Cannelli}}, \bibinfo {author} {\bibfnamefont {J.}~\bibnamefont {Wiktor}},
  \bibinfo {author} {\bibfnamefont {N.}~\bibnamefont {Colonna}}, \bibinfo
  {author} {\bibfnamefont {L.}~\bibnamefont {Leroy}}, \bibinfo {author}
  {\bibfnamefont {M.}~\bibnamefont {Puppin}}, \bibinfo {author} {\bibfnamefont
  {C.}~\bibnamefont {Bacellar}}, \bibinfo {author} {\bibfnamefont
  {I.}~\bibnamefont {Sadykov}}, \bibinfo {author} {\bibfnamefont
  {F.}~\bibnamefont {Krieg}}, \bibinfo {author} {\bibfnamefont
  {G.}~\bibnamefont {Smolentsev}}, \bibinfo {author} {\bibfnamefont {M.~V.}\
  \bibnamefont {Kovalenko}}, \emph {et~al.},\ }\href
  {https://doi.org/10.1021/acs.jpclett.2c00281} {\bibfield  {journal} {\bibinfo
   {journal} {The Journal of Physical Chemistry Letters}\ }\textbf {\bibinfo
  {volume} {13}},\ \bibinfo {pages} {3382} (\bibinfo {year}
  {2022})}\BibitemShut {NoStop}%
\bibitem [{\citenamefont {Jinnouchi}\ \emph {et~al.}(2019)\citenamefont
  {Jinnouchi}, \citenamefont {Lahnsteiner}, \citenamefont {Karsai},
  \citenamefont {Kresse},\ and\ \citenamefont {Bokdam}}]{jinnouchi2019phase}%
  \BibitemOpen
  \bibfield  {author} {\bibinfo {author} {\bibfnamefont {R.}~\bibnamefont
  {Jinnouchi}}, \bibinfo {author} {\bibfnamefont {J.}~\bibnamefont
  {Lahnsteiner}}, \bibinfo {author} {\bibfnamefont {F.}~\bibnamefont {Karsai}},
  \bibinfo {author} {\bibfnamefont {G.}~\bibnamefont {Kresse}},\ and\ \bibinfo
  {author} {\bibfnamefont {M.}~\bibnamefont {Bokdam}},\ }\href
  {https://doi.org/10.1103/PhysRevLett.122.225701} {\bibfield  {journal}
  {\bibinfo  {journal} {Physical Review Letters}\ }\textbf {\bibinfo {volume}
  {122}},\ \bibinfo {pages} {225701} (\bibinfo {year} {2019})}\BibitemShut
  {NoStop}%
\bibitem [{\citenamefont {Lahnsteiner}\ \emph {et~al.}(2019)\citenamefont
  {Lahnsteiner}, \citenamefont {Jinnouchi},\ and\ \citenamefont
  {Bokdam}}]{lahnsteiner2019long}%
  \BibitemOpen
  \bibfield  {author} {\bibinfo {author} {\bibfnamefont {J.}~\bibnamefont
  {Lahnsteiner}}, \bibinfo {author} {\bibfnamefont {R.}~\bibnamefont
  {Jinnouchi}},\ and\ \bibinfo {author} {\bibfnamefont {M.}~\bibnamefont
  {Bokdam}},\ }\href {https://doi.org/10.1103/PhysRevB.100.094106} {\bibfield
  {journal} {\bibinfo  {journal} {Physical Review B}\ }\textbf {\bibinfo
  {volume} {100}},\ \bibinfo {pages} {094106} (\bibinfo {year}
  {2019})}\BibitemShut {NoStop}%
\bibitem [{\citenamefont {Thomas}\ \emph {et~al.}(2019)\citenamefont {Thomas},
  \citenamefont {Bechtel}, \citenamefont {Natarajan},\ and\ \citenamefont
  {Van~der Ven}}]{thomas2019machine}%
  \BibitemOpen
  \bibfield  {author} {\bibinfo {author} {\bibfnamefont {J.~C.}\ \bibnamefont
  {Thomas}}, \bibinfo {author} {\bibfnamefont {J.~S.}\ \bibnamefont {Bechtel}},
  \bibinfo {author} {\bibfnamefont {A.~R.}\ \bibnamefont {Natarajan}},\ and\
  \bibinfo {author} {\bibfnamefont {A.}~\bibnamefont {Van~der Ven}},\ }\href
  {https://doi.org/10.1103/PhysRevB.100.134101} {\bibfield  {journal} {\bibinfo
   {journal} {Physical Review B}\ }\textbf {\bibinfo {volume} {100}},\ \bibinfo
  {pages} {134101} (\bibinfo {year} {2019})}\BibitemShut {NoStop}%
\bibitem [{\citenamefont {Zhou}\ \emph {et~al.}(2020)\citenamefont {Zhou},
  \citenamefont {Chu},\ and\ \citenamefont {Prezhdo}}]{zhou2020structural}%
  \BibitemOpen
  \bibfield  {author} {\bibinfo {author} {\bibfnamefont {G.}~\bibnamefont
  {Zhou}}, \bibinfo {author} {\bibfnamefont {W.}~\bibnamefont {Chu}},\ and\
  \bibinfo {author} {\bibfnamefont {O.~V.}\ \bibnamefont {Prezhdo}},\ }\href
  {https://doi.org/10.1021/acsenergylett.0c00899} {\bibfield  {journal}
  {\bibinfo  {journal} {ACS Energy Letters}\ }\textbf {\bibinfo {volume} {5}},\
  \bibinfo {pages} {1930} (\bibinfo {year} {2020})}\BibitemShut {NoStop}%
\bibitem [{\citenamefont {Mangan}\ \emph {et~al.}(2021)\citenamefont {Mangan},
  \citenamefont {Zhou}, \citenamefont {Chu},\ and\ \citenamefont
  {Prezhdo}}]{mangan2021dependence}%
  \BibitemOpen
  \bibfield  {author} {\bibinfo {author} {\bibfnamefont {S.~M.}\ \bibnamefont
  {Mangan}}, \bibinfo {author} {\bibfnamefont {G.}~\bibnamefont {Zhou}},
  \bibinfo {author} {\bibfnamefont {W.}~\bibnamefont {Chu}},\ and\ \bibinfo
  {author} {\bibfnamefont {O.~V.}\ \bibnamefont {Prezhdo}},\ }\href
  {https://doi.org/10.1021/acs.jpclett.1c02361} {\bibfield  {journal} {\bibinfo
   {journal} {The Journal of Physical Chemistry Letters}\ }\textbf {\bibinfo
  {volume} {12}},\ \bibinfo {pages} {8672} (\bibinfo {year}
  {2021})}\BibitemShut {NoStop}%
\bibitem [{\citenamefont {Bokdam}\ \emph {et~al.}(2021)\citenamefont {Bokdam},
  \citenamefont {Lahnsteiner},\ and\ \citenamefont
  {Sarma}}]{bokdam2021exploring}%
  \BibitemOpen
  \bibfield  {author} {\bibinfo {author} {\bibfnamefont {M.}~\bibnamefont
  {Bokdam}}, \bibinfo {author} {\bibfnamefont {J.}~\bibnamefont
  {Lahnsteiner}},\ and\ \bibinfo {author} {\bibfnamefont {D.~D.}\ \bibnamefont
  {Sarma}},\ }\href {https://doi.org/10.1021/acs.jpcc.1c06835} {\bibfield
  {journal} {\bibinfo  {journal} {The Journal of Physical Chemistry C}\
  }\textbf {\bibinfo {volume} {125}},\ \bibinfo {pages} {21077} (\bibinfo
  {year} {2021})}\BibitemShut {NoStop}%
\bibitem [{\citenamefont {Gr\"{u}ninger}\ \emph {et~al.}(2021)\citenamefont
  {Gr\"{u}ninger}, \citenamefont {Bokdam}, \citenamefont {Leupold},
  \citenamefont {Tinnemans}, \citenamefont {Moos}, \citenamefont {Wijs},
  \citenamefont {Panzer},\ and\ \citenamefont
  {Kentgens}}]{gruninger2021microscopic}%
  \BibitemOpen
  \bibfield  {author} {\bibinfo {author} {\bibfnamefont {H.}~\bibnamefont
  {Gr\"{u}ninger}}, \bibinfo {author} {\bibfnamefont {M.}~\bibnamefont
  {Bokdam}}, \bibinfo {author} {\bibfnamefont {N.}~\bibnamefont {Leupold}},
  \bibinfo {author} {\bibfnamefont {P.}~\bibnamefont {Tinnemans}}, \bibinfo
  {author} {\bibfnamefont {R.}~\bibnamefont {Moos}}, \bibinfo {author}
  {\bibfnamefont {G.~A.~D.}\ \bibnamefont {Wijs}}, \bibinfo {author}
  {\bibfnamefont {F.}~\bibnamefont {Panzer}},\ and\ \bibinfo {author}
  {\bibfnamefont {A.~P.~M.}\ \bibnamefont {Kentgens}},\ }\href
  {https://doi.org/10.1021/acs.jpcc.0c10042} {\bibfield  {journal} {\bibinfo
  {journal} {The Journal of Physical Chemistry C}\ }\textbf {\bibinfo {volume}
  {125}},\ \bibinfo {pages} {1742} (\bibinfo {year} {2021})}\BibitemShut
  {NoStop}%
\bibitem [{\citenamefont {Lahnsteiner}\ and\ \citenamefont
  {Bokdam}(2022)}]{LahBok22}%
  \BibitemOpen
  \bibfield  {author} {\bibinfo {author} {\bibfnamefont {J.}~\bibnamefont
  {Lahnsteiner}}\ and\ \bibinfo {author} {\bibfnamefont {M.}~\bibnamefont
  {Bokdam}},\ }\href {https://doi.org/10.1103/PhysRevB.105.024302} {\bibfield
  {journal} {\bibinfo  {journal} {Physical Review B}\ }\textbf {\bibinfo
  {volume} {105}},\ \bibinfo {pages} {024302} (\bibinfo {year}
  {2022})}\BibitemShut {NoStop}%
\bibitem [{\citenamefont {Braeckevelt}\ \emph {et~al.}(2022)\citenamefont
  {Braeckevelt}, \citenamefont {Goeminne}, \citenamefont {Vandenhaute},
  \citenamefont {Borgmans}, \citenamefont {Verstraelen}, \citenamefont
  {Steele}, \citenamefont {Roeffaers}, \citenamefont {Hofkens}, \citenamefont
  {Rogge},\ and\ \citenamefont {Van~Speybroeck}}]{BraGoeVan22}%
  \BibitemOpen
  \bibfield  {author} {\bibinfo {author} {\bibfnamefont {T.}~\bibnamefont
  {Braeckevelt}}, \bibinfo {author} {\bibfnamefont {R.}~\bibnamefont
  {Goeminne}}, \bibinfo {author} {\bibfnamefont {S.}~\bibnamefont
  {Vandenhaute}}, \bibinfo {author} {\bibfnamefont {S.}~\bibnamefont
  {Borgmans}}, \bibinfo {author} {\bibfnamefont {T.}~\bibnamefont
  {Verstraelen}}, \bibinfo {author} {\bibfnamefont {J.~A.}\ \bibnamefont
  {Steele}}, \bibinfo {author} {\bibfnamefont {M.~B.~J.}\ \bibnamefont
  {Roeffaers}}, \bibinfo {author} {\bibfnamefont {J.}~\bibnamefont {Hofkens}},
  \bibinfo {author} {\bibfnamefont {S.~M.~J.}\ \bibnamefont {Rogge}},\ and\
  \bibinfo {author} {\bibfnamefont {V.}~\bibnamefont {Van~Speybroeck}},\ }\href
  {https://doi.org/10.1021/acs.chemmater.2c01508} {\bibfield  {journal}
  {\bibinfo  {journal} {Chemistry of Materials}\ }\textbf {\bibinfo {volume}
  {34}},\ \bibinfo {pages} {8561} (\bibinfo {year} {2022})}\BibitemShut
  {NoStop}%
\bibitem [{\citenamefont {Fransson}\ \emph {et~al.}(2022)\citenamefont
  {Fransson}, \citenamefont {Rosander}, \citenamefont {Eriksson}, \citenamefont
  {Rahm}, \citenamefont {Tadano},\ and\ \citenamefont {Erhart}}]{FraRosEri22}%
  \BibitemOpen
  \bibfield  {author} {\bibinfo {author} {\bibfnamefont {E.}~\bibnamefont
  {Fransson}}, \bibinfo {author} {\bibfnamefont {P.}~\bibnamefont {Rosander}},
  \bibinfo {author} {\bibfnamefont {F.}~\bibnamefont {Eriksson}}, \bibinfo
  {author} {\bibfnamefont {J.~M.}\ \bibnamefont {Rahm}}, \bibinfo {author}
  {\bibfnamefont {T.}~\bibnamefont {Tadano}},\ and\ \bibinfo {author}
  {\bibfnamefont {P.}~\bibnamefont {Erhart}},\ }\href
  {https://doi.org/10.48550/arXiv.2211.08197} {\bibinfo {title} {Probing the
  limits of the phonon quasi-particle picture: {{The}} transition from
  underdamped to overdamped dynamics in {{CsPbBr3}}}} (\bibinfo {year}
  {2022}),\ \Eprint {https://arxiv.org/abs/2211.08197} {arXiv:2211.08197
  [cond-mat]} \BibitemShut {NoStop}%
\bibitem [{\citenamefont {Ghosh}\ \emph {et~al.}(2017)\citenamefont {Ghosh},
  \citenamefont {Walsh~Atkins}, \citenamefont {Islam}, \citenamefont {Walker},\
  and\ \citenamefont {Eames}}]{ghosh2017good}%
  \BibitemOpen
  \bibfield  {author} {\bibinfo {author} {\bibfnamefont {D.}~\bibnamefont
  {Ghosh}}, \bibinfo {author} {\bibfnamefont {P.}~\bibnamefont {Walsh~Atkins}},
  \bibinfo {author} {\bibfnamefont {M.~S.}\ \bibnamefont {Islam}}, \bibinfo
  {author} {\bibfnamefont {A.~B.}\ \bibnamefont {Walker}},\ and\ \bibinfo
  {author} {\bibfnamefont {C.}~\bibnamefont {Eames}},\ }\href
  {https://doi.org/10.1021/acsenergylett.7b00729} {\bibfield  {journal}
  {\bibinfo  {journal} {ACS Energy Letters}\ }\textbf {\bibinfo {volume} {2}},\
  \bibinfo {pages} {2424} (\bibinfo {year} {2017})}\BibitemShut {NoStop}%
\bibitem [{\citenamefont {Sun}\ \emph {et~al.}(2015)\citenamefont {Sun},
  \citenamefont {Ruzsinszky},\ and\ \citenamefont {Perdew}}]{SunRuzPer15}%
  \BibitemOpen
  \bibfield  {author} {\bibinfo {author} {\bibfnamefont {J.}~\bibnamefont
  {Sun}}, \bibinfo {author} {\bibfnamefont {A.}~\bibnamefont {Ruzsinszky}},\
  and\ \bibinfo {author} {\bibfnamefont {J.~P.}\ \bibnamefont {Perdew}},\
  }\href {https://doi.org/10.1103/PhysRevLett.115.036402} {\bibfield  {journal}
  {\bibinfo  {journal} {Physical Review Letters}\ }\textbf {\bibinfo {volume}
  {115}},\ \bibinfo {pages} {036402} (\bibinfo {year} {2015})}\BibitemShut
  {NoStop}%
\bibitem [{\citenamefont {Berland}\ and\ \citenamefont
  {Hyldgaard}(2014)}]{BerHyl2014}%
  \BibitemOpen
  \bibfield  {author} {\bibinfo {author} {\bibfnamefont {K.}~\bibnamefont
  {Berland}}\ and\ \bibinfo {author} {\bibfnamefont {P.}~\bibnamefont
  {Hyldgaard}},\ }\href {https://doi.org/10.1103/PhysRevB.89.035412} {\bibfield
   {journal} {\bibinfo  {journal} {Physical Review B}\ }\textbf {\bibinfo
  {volume} {89}},\ \bibinfo {pages} {035412} (\bibinfo {year}
  {2014})}\BibitemShut {NoStop}%
\bibitem [{\citenamefont {Perdew}\ \emph {et~al.}(2008)\citenamefont {Perdew},
  \citenamefont {Ruzsinszky}, \citenamefont {Csonka}, \citenamefont {Vydrov},
  \citenamefont {Scuseria}, \citenamefont {Constantin}, \citenamefont {Zhou},\
  and\ \citenamefont {Burke}}]{PerRuz2008}%
  \BibitemOpen
  \bibfield  {author} {\bibinfo {author} {\bibfnamefont {J.~P.}\ \bibnamefont
  {Perdew}}, \bibinfo {author} {\bibfnamefont {A.}~\bibnamefont {Ruzsinszky}},
  \bibinfo {author} {\bibfnamefont {G.~I.}\ \bibnamefont {Csonka}}, \bibinfo
  {author} {\bibfnamefont {O.~A.}\ \bibnamefont {Vydrov}}, \bibinfo {author}
  {\bibfnamefont {G.~E.}\ \bibnamefont {Scuseria}}, \bibinfo {author}
  {\bibfnamefont {L.~A.}\ \bibnamefont {Constantin}}, \bibinfo {author}
  {\bibfnamefont {X.}~\bibnamefont {Zhou}},\ and\ \bibinfo {author}
  {\bibfnamefont {K.}~\bibnamefont {Burke}},\ }\href
  {https://doi.org/10.1103/PhysRevLett.100.136406} {\bibfield  {journal}
  {\bibinfo  {journal} {Physical Review Letters}\ }\textbf {\bibinfo {volume}
  {100}},\ \bibinfo {pages} {136406} (\bibinfo {year} {2008})}\BibitemShut
  {NoStop}%
\bibitem [{\citenamefont {Perdew}\ \emph {et~al.}(1996)\citenamefont {Perdew},
  \citenamefont {Burke},\ and\ \citenamefont {Ernzerhof}}]{PerBurErn}%
  \BibitemOpen
  \bibfield  {author} {\bibinfo {author} {\bibfnamefont {J.~P.}\ \bibnamefont
  {Perdew}}, \bibinfo {author} {\bibfnamefont {K.}~\bibnamefont {Burke}},\ and\
  \bibinfo {author} {\bibfnamefont {M.}~\bibnamefont {Ernzerhof}},\ }\href
  {https://doi.org/10.1103/PhysRevLett.77.3865} {\bibfield  {journal} {\bibinfo
   {journal} {Physical Review Letters}\ }\textbf {\bibinfo {volume} {77}},\
  \bibinfo {pages} {3865} (\bibinfo {year} {1996})}\BibitemShut {NoStop}%
\bibitem [{\citenamefont {Hirotsu}\ \emph {et~al.}(1974)\citenamefont
  {Hirotsu}, \citenamefont {Harada}, \citenamefont {Iizumi},\ and\
  \citenamefont {Gesi}}]{Hirotsu1974}%
  \BibitemOpen
  \bibfield  {author} {\bibinfo {author} {\bibfnamefont {S.}~\bibnamefont
  {Hirotsu}}, \bibinfo {author} {\bibfnamefont {J.}~\bibnamefont {Harada}},
  \bibinfo {author} {\bibfnamefont {M.}~\bibnamefont {Iizumi}},\ and\ \bibinfo
  {author} {\bibfnamefont {K.}~\bibnamefont {Gesi}},\ }\href
  {https://doi.org/10.1143/JPSJ.37.1393} {\bibfield  {journal} {\bibinfo
  {journal} {Journal of the Physical Society of Japan}\ }\textbf {\bibinfo
  {volume} {37}},\ \bibinfo {pages} {1393} (\bibinfo {year}
  {1974})}\BibitemShut {NoStop}%
\bibitem [{\citenamefont {Rodov{\'{a}}}\ \emph {et~al.}(2003)\citenamefont
  {Rodov{\'{a}}}, \citenamefont {Bro{\v{z}}ek}, \citenamefont
  {Kn{\'{\i}}{\v{z}}ek},\ and\ \citenamefont {Nitsch}}]{Rodov2003}%
  \BibitemOpen
  \bibfield  {author} {\bibinfo {author} {\bibfnamefont {M.}~\bibnamefont
  {Rodov{\'{a}}}}, \bibinfo {author} {\bibfnamefont {J.}~\bibnamefont
  {Bro{\v{z}}ek}}, \bibinfo {author} {\bibfnamefont {K.}~\bibnamefont
  {Kn{\'{\i}}{\v{z}}ek}},\ and\ \bibinfo {author} {\bibfnamefont
  {K.}~\bibnamefont {Nitsch}},\ }\href
  {https://doi.org/10.1023/a:1022836800820} {\bibfield  {journal} {\bibinfo
  {journal} {Journal of Thermal Analysis and Calorimetry}\ }\textbf {\bibinfo
  {volume} {71}},\ \bibinfo {pages} {667} (\bibinfo {year} {2003})}\BibitemShut
  {NoStop}%
\bibitem [{\citenamefont {Malyshkin}\ \emph {et~al.}(2020)\citenamefont
  {Malyshkin}, \citenamefont {Sereda}, \citenamefont {Ivanov}, \citenamefont
  {Mazurin}, \citenamefont {Sednev-Lugovets}, \citenamefont {Tsvetkov},\ and\
  \citenamefont {Zuev}}]{Malyshkin2020}%
  \BibitemOpen
  \bibfield  {author} {\bibinfo {author} {\bibfnamefont {D.}~\bibnamefont
  {Malyshkin}}, \bibinfo {author} {\bibfnamefont {V.}~\bibnamefont {Sereda}},
  \bibinfo {author} {\bibfnamefont {I.}~\bibnamefont {Ivanov}}, \bibinfo
  {author} {\bibfnamefont {M.}~\bibnamefont {Mazurin}}, \bibinfo {author}
  {\bibfnamefont {A.}~\bibnamefont {Sednev-Lugovets}}, \bibinfo {author}
  {\bibfnamefont {D.}~\bibnamefont {Tsvetkov}},\ and\ \bibinfo {author}
  {\bibfnamefont {A.}~\bibnamefont {Zuev}},\ }\href
  {https://doi.org/10.1016/j.matlet.2020.128458} {\bibfield  {journal}
  {\bibinfo  {journal} {Materials Letters}\ }\textbf {\bibinfo {volume}
  {278}},\ \bibinfo {pages} {128458} (\bibinfo {year} {2020})}\BibitemShut
  {NoStop}%
\bibitem [{\citenamefont {Klarbring}(2019)}]{Klarbring2019}%
  \BibitemOpen
  \bibfield  {author} {\bibinfo {author} {\bibfnamefont {J.}~\bibnamefont
  {Klarbring}},\ }\href {https://doi.org/10.1103/physrevb.99.104105} {\bibfield
   {journal} {\bibinfo  {journal} {Physical Review B}\ }\textbf {\bibinfo
  {volume} {99}},\ \bibinfo {pages} {104105} (\bibinfo {year}
  {2019})}\BibitemShut {NoStop}%
\bibitem [{\citenamefont {Stoumpos}\ \emph {et~al.}(2013)\citenamefont
  {Stoumpos}, \citenamefont {Malliakas}, \citenamefont {Peters}, \citenamefont
  {Liu}, \citenamefont {Sebastian}, \citenamefont {Im}, \citenamefont
  {Chasapis}, \citenamefont {Wibowo}, \citenamefont {Chung}, \citenamefont
  {Freeman}, \citenamefont {Wessels},\ and\ \citenamefont
  {Kanatzidis}}]{Stoumpos2013}%
  \BibitemOpen
  \bibfield  {author} {\bibinfo {author} {\bibfnamefont {C.~C.}\ \bibnamefont
  {Stoumpos}}, \bibinfo {author} {\bibfnamefont {C.~D.}\ \bibnamefont
  {Malliakas}}, \bibinfo {author} {\bibfnamefont {J.~A.}\ \bibnamefont
  {Peters}}, \bibinfo {author} {\bibfnamefont {Z.}~\bibnamefont {Liu}},
  \bibinfo {author} {\bibfnamefont {M.}~\bibnamefont {Sebastian}}, \bibinfo
  {author} {\bibfnamefont {J.}~\bibnamefont {Im}}, \bibinfo {author}
  {\bibfnamefont {T.~C.}\ \bibnamefont {Chasapis}}, \bibinfo {author}
  {\bibfnamefont {A.~C.}\ \bibnamefont {Wibowo}}, \bibinfo {author}
  {\bibfnamefont {D.~Y.}\ \bibnamefont {Chung}}, \bibinfo {author}
  {\bibfnamefont {A.~J.}\ \bibnamefont {Freeman}}, \bibinfo {author}
  {\bibfnamefont {B.~W.}\ \bibnamefont {Wessels}},\ and\ \bibinfo {author}
  {\bibfnamefont {M.~G.}\ \bibnamefont {Kanatzidis}},\ }\href
  {https://doi.org/10.1021/cg400645t} {\bibfield  {journal} {\bibinfo
  {journal} {Crystal Growth {\&} Design}\ }\textbf {\bibinfo {volume} {13}},\
  \bibinfo {pages} {2722} (\bibinfo {year} {2013})}\BibitemShut {NoStop}%
\bibitem [{\citenamefont {He}\ \emph {et~al.}(2021)\citenamefont {He},
  \citenamefont {Stoumpos}, \citenamefont {Hadar}, \citenamefont {Luo},
  \citenamefont {McCall}, \citenamefont {Liu}, \citenamefont {Chung},
  \citenamefont {Wessels},\ and\ \citenamefont {Kanatzidis}}]{HeStoHad21}%
  \BibitemOpen
  \bibfield  {author} {\bibinfo {author} {\bibfnamefont {Y.}~\bibnamefont
  {He}}, \bibinfo {author} {\bibfnamefont {C.~C.}\ \bibnamefont {Stoumpos}},
  \bibinfo {author} {\bibfnamefont {I.}~\bibnamefont {Hadar}}, \bibinfo
  {author} {\bibfnamefont {Z.}~\bibnamefont {Luo}}, \bibinfo {author}
  {\bibfnamefont {K.~M.}\ \bibnamefont {McCall}}, \bibinfo {author}
  {\bibfnamefont {Z.}~\bibnamefont {Liu}}, \bibinfo {author} {\bibfnamefont
  {D.~Y.}\ \bibnamefont {Chung}}, \bibinfo {author} {\bibfnamefont {B.~W.}\
  \bibnamefont {Wessels}},\ and\ \bibinfo {author} {\bibfnamefont {M.~G.}\
  \bibnamefont {Kanatzidis}},\ }\href {https://doi.org/10.1021/jacs.0c12254}
  {\bibfield  {journal} {\bibinfo  {journal} {Journal of the American Chemical
  Society}\ }\textbf {\bibinfo {volume} {143}},\ \bibinfo {pages} {2068}
  (\bibinfo {year} {2021})}\BibitemShut {NoStop}%
\bibitem [{\citenamefont {Gharaee}\ \emph {et~al.}(2017)\citenamefont
  {Gharaee}, \citenamefont {Erhart},\ and\ \citenamefont
  {Hyldgaard}}]{GhaErhHyl17}%
  \BibitemOpen
  \bibfield  {author} {\bibinfo {author} {\bibfnamefont {L.}~\bibnamefont
  {Gharaee}}, \bibinfo {author} {\bibfnamefont {P.}~\bibnamefont {Erhart}},\
  and\ \bibinfo {author} {\bibfnamefont {P.}~\bibnamefont {Hyldgaard}},\ }\href
  {https://doi.org/10.1103/PhysRevB.95.085147} {\bibfield  {journal} {\bibinfo
  {journal} {Physical Review B}\ }\textbf {\bibinfo {volume} {95}},\ \bibinfo
  {pages} {085147} (\bibinfo {year} {2017})}\BibitemShut {NoStop}%
\bibitem [{\citenamefont {Sutton}\ \emph {et~al.}(2018)\citenamefont {Sutton},
  \citenamefont {Filip}, \citenamefont {Haghighirad}, \citenamefont {Sakai},
  \citenamefont {Wenger}, \citenamefont {Giustino},\ and\ \citenamefont
  {Snaith}}]{SutFilHag18}%
  \BibitemOpen
  \bibfield  {author} {\bibinfo {author} {\bibfnamefont {R.~J.}\ \bibnamefont
  {Sutton}}, \bibinfo {author} {\bibfnamefont {M.~R.}\ \bibnamefont {Filip}},
  \bibinfo {author} {\bibfnamefont {A.~A.}\ \bibnamefont {Haghighirad}},
  \bibinfo {author} {\bibfnamefont {N.}~\bibnamefont {Sakai}}, \bibinfo
  {author} {\bibfnamefont {B.}~\bibnamefont {Wenger}}, \bibinfo {author}
  {\bibfnamefont {F.}~\bibnamefont {Giustino}},\ and\ \bibinfo {author}
  {\bibfnamefont {H.~J.}\ \bibnamefont {Snaith}},\ }\href
  {https://doi.org/10.1021/acsenergylett.8b00672} {\bibfield  {journal}
  {\bibinfo  {journal} {ACS Energy Letters}\ }\textbf {\bibinfo {volume} {3}},\
  \bibinfo {pages} {1787} (\bibinfo {year} {2018})}\BibitemShut {NoStop}%
\bibitem [{\citenamefont {Dastidar}\ \emph {et~al.}(2017)\citenamefont
  {Dastidar}, \citenamefont {Hawley}, \citenamefont {Dillon}, \citenamefont
  {{Gutierrez-Perez}}, \citenamefont {Spanier},\ and\ \citenamefont
  {Fafarman}}]{DasHawDil17}%
  \BibitemOpen
  \bibfield  {author} {\bibinfo {author} {\bibfnamefont {S.}~\bibnamefont
  {Dastidar}}, \bibinfo {author} {\bibfnamefont {C.~J.}\ \bibnamefont
  {Hawley}}, \bibinfo {author} {\bibfnamefont {A.~D.}\ \bibnamefont {Dillon}},
  \bibinfo {author} {\bibfnamefont {A.~D.}\ \bibnamefont {{Gutierrez-Perez}}},
  \bibinfo {author} {\bibfnamefont {J.~E.}\ \bibnamefont {Spanier}},\ and\
  \bibinfo {author} {\bibfnamefont {A.~T.}\ \bibnamefont {Fafarman}},\ }\href
  {https://doi.org/10.1021/acs.jpclett.7b00134} {\bibfield  {journal} {\bibinfo
   {journal} {The Journal of Physical Chemistry Letters}\ }\textbf {\bibinfo
  {volume} {8}},\ \bibinfo {pages} {1278} (\bibinfo {year} {2017})}\BibitemShut
  {NoStop}%
\bibitem [{\citenamefont {Marronnier}\ \emph
  {et~al.}(2018{\natexlab{b}})\citenamefont {Marronnier}, \citenamefont {Roma},
  \citenamefont {{Boyer-Richard}}, \citenamefont {Pedesseau}, \citenamefont
  {Jancu}, \citenamefont {Bonnassieux}, \citenamefont {Katan}, \citenamefont
  {Stoumpos}, \citenamefont {Kanatzidis},\ and\ \citenamefont
  {Even}}]{MarRomBoy18}%
  \BibitemOpen
  \bibfield  {author} {\bibinfo {author} {\bibfnamefont {A.}~\bibnamefont
  {Marronnier}}, \bibinfo {author} {\bibfnamefont {G.}~\bibnamefont {Roma}},
  \bibinfo {author} {\bibfnamefont {S.}~\bibnamefont {{Boyer-Richard}}},
  \bibinfo {author} {\bibfnamefont {L.}~\bibnamefont {Pedesseau}}, \bibinfo
  {author} {\bibfnamefont {J.-M.}\ \bibnamefont {Jancu}}, \bibinfo {author}
  {\bibfnamefont {Y.}~\bibnamefont {Bonnassieux}}, \bibinfo {author}
  {\bibfnamefont {C.}~\bibnamefont {Katan}}, \bibinfo {author} {\bibfnamefont
  {C.~C.}\ \bibnamefont {Stoumpos}}, \bibinfo {author} {\bibfnamefont {M.~G.}\
  \bibnamefont {Kanatzidis}},\ and\ \bibinfo {author} {\bibfnamefont
  {J.}~\bibnamefont {Even}},\ }\href {https://doi.org/10.1021/acsnano.8b00267}
  {\bibfield  {journal} {\bibinfo  {journal} {ACS Nano}\ }\textbf {\bibinfo
  {volume} {12}},\ \bibinfo {pages} {3477} (\bibinfo {year}
  {2018}{\natexlab{b}})}\BibitemShut {NoStop}%
\bibitem [{\citenamefont {Ke}\ \emph {et~al.}(2021)\citenamefont {Ke},
  \citenamefont {Wang}, \citenamefont {Jia}, \citenamefont {Wolf},
  \citenamefont {Yan}, \citenamefont {Niu}, \citenamefont {Devereaux},
  \citenamefont {Karunadasa}, \citenamefont {Mao},\ and\ \citenamefont
  {Lin}}]{KeWanJia21}%
  \BibitemOpen
  \bibfield  {author} {\bibinfo {author} {\bibfnamefont {F.}~\bibnamefont
  {Ke}}, \bibinfo {author} {\bibfnamefont {C.}~\bibnamefont {Wang}}, \bibinfo
  {author} {\bibfnamefont {C.}~\bibnamefont {Jia}}, \bibinfo {author}
  {\bibfnamefont {N.~R.}\ \bibnamefont {Wolf}}, \bibinfo {author}
  {\bibfnamefont {J.}~\bibnamefont {Yan}}, \bibinfo {author} {\bibfnamefont
  {S.}~\bibnamefont {Niu}}, \bibinfo {author} {\bibfnamefont {T.~P.}\
  \bibnamefont {Devereaux}}, \bibinfo {author} {\bibfnamefont {H.~I.}\
  \bibnamefont {Karunadasa}}, \bibinfo {author} {\bibfnamefont {W.~L.}\
  \bibnamefont {Mao}},\ and\ \bibinfo {author} {\bibfnamefont {Y.}~\bibnamefont
  {Lin}},\ }\href {https://doi.org/10.1038/s41467-020-20745-5} {\bibfield
  {journal} {\bibinfo  {journal} {Nature Communications}\ }\textbf {\bibinfo
  {volume} {12}},\ \bibinfo {pages} {461} (\bibinfo {year} {2021})}\BibitemShut
  {NoStop}%
\bibitem [{\citenamefont {Paul}\ \emph {et~al.}(2017)\citenamefont {Paul},
  \citenamefont {Sun}, \citenamefont {Perdew},\ and\ \citenamefont
  {Waghmare}}]{PauSunPer17}%
  \BibitemOpen
  \bibfield  {author} {\bibinfo {author} {\bibfnamefont {A.}~\bibnamefont
  {Paul}}, \bibinfo {author} {\bibfnamefont {J.}~\bibnamefont {Sun}}, \bibinfo
  {author} {\bibfnamefont {J.~P.}\ \bibnamefont {Perdew}},\ and\ \bibinfo
  {author} {\bibfnamefont {U.~V.}\ \bibnamefont {Waghmare}},\ }\href
  {https://doi.org/10.1103/PhysRevB.95.054111} {\bibfield  {journal} {\bibinfo
  {journal} {Physical Review B}\ }\textbf {\bibinfo {volume} {95}},\ \bibinfo
  {pages} {054111} (\bibinfo {year} {2017})}\BibitemShut {NoStop}%
\bibitem [{\citenamefont {Zhang}\ \emph {et~al.}(2017)\citenamefont {Zhang},
  \citenamefont {Sun}, \citenamefont {Perdew},\ and\ \citenamefont
  {Wu}}]{ZhaSunPer17}%
  \BibitemOpen
  \bibfield  {author} {\bibinfo {author} {\bibfnamefont {Y.}~\bibnamefont
  {Zhang}}, \bibinfo {author} {\bibfnamefont {J.}~\bibnamefont {Sun}}, \bibinfo
  {author} {\bibfnamefont {J.~P.}\ \bibnamefont {Perdew}},\ and\ \bibinfo
  {author} {\bibfnamefont {X.}~\bibnamefont {Wu}},\ }\href
  {https://doi.org/10.1103/PhysRevB.96.035143} {\bibfield  {journal} {\bibinfo
  {journal} {Physical Review B}\ }\textbf {\bibinfo {volume} {96}},\ \bibinfo
  {pages} {035143} (\bibinfo {year} {2017})}\BibitemShut {NoStop}%
\bibitem [{\citenamefont {Verdi}\ \emph {et~al.}(2022)\citenamefont {Verdi},
  \citenamefont {Ranalli}, \citenamefont {Franchini},\ and\ \citenamefont
  {Kresse}}]{VerRanFra22}%
  \BibitemOpen
  \bibfield  {author} {\bibinfo {author} {\bibfnamefont {C.}~\bibnamefont
  {Verdi}}, \bibinfo {author} {\bibfnamefont {L.}~\bibnamefont {Ranalli}},
  \bibinfo {author} {\bibfnamefont {C.}~\bibnamefont {Franchini}},\ and\
  \bibinfo {author} {\bibfnamefont {G.}~\bibnamefont {Kresse}},\ }\href
  {https://doi.org/10.48550/arXiv.2211.09616} {\bibinfo {title} {Quantum
  paraelectricity and structural phase transitions in strontium titanate beyond
  density-functional theory}} (\bibinfo {year} {2022}),\ \Eprint
  {https://arxiv.org/abs/2211.09616} {arXiv:2211.09616 [cond-mat]} \BibitemShut
  {NoStop}%
\bibitem [{\citenamefont {Fan}\ \emph {et~al.}(2022)\citenamefont {Fan},
  \citenamefont {Wang}, \citenamefont {Ying}, \citenamefont {Song},
  \citenamefont {Wang}, \citenamefont {Wang}, \citenamefont {Zeng},
  \citenamefont {Xu}, \citenamefont {Lindgren}, \citenamefont {Rahm},
  \citenamefont {Gabourie}, \citenamefont {Liu}, \citenamefont {Dong},
  \citenamefont {Wu}, \citenamefont {Chen}, \citenamefont {Zhong},
  \citenamefont {Sun}, \citenamefont {Erhart}, \citenamefont {Su},\ and\
  \citenamefont {Ala-Nissila}}]{FanWanYin22}%
  \BibitemOpen
  \bibfield  {author} {\bibinfo {author} {\bibfnamefont {Z.}~\bibnamefont
  {Fan}}, \bibinfo {author} {\bibfnamefont {Y.}~\bibnamefont {Wang}}, \bibinfo
  {author} {\bibfnamefont {P.}~\bibnamefont {Ying}}, \bibinfo {author}
  {\bibfnamefont {K.}~\bibnamefont {Song}}, \bibinfo {author} {\bibfnamefont
  {J.}~\bibnamefont {Wang}}, \bibinfo {author} {\bibfnamefont {Y.}~\bibnamefont
  {Wang}}, \bibinfo {author} {\bibfnamefont {Z.}~\bibnamefont {Zeng}}, \bibinfo
  {author} {\bibfnamefont {K.}~\bibnamefont {Xu}}, \bibinfo {author}
  {\bibfnamefont {E.}~\bibnamefont {Lindgren}}, \bibinfo {author}
  {\bibfnamefont {J.~M.}\ \bibnamefont {Rahm}}, \bibinfo {author}
  {\bibfnamefont {A.~J.}\ \bibnamefont {Gabourie}}, \bibinfo {author}
  {\bibfnamefont {J.}~\bibnamefont {Liu}}, \bibinfo {author} {\bibfnamefont
  {H.}~\bibnamefont {Dong}}, \bibinfo {author} {\bibfnamefont {J.}~\bibnamefont
  {Wu}}, \bibinfo {author} {\bibfnamefont {Y.}~\bibnamefont {Chen}}, \bibinfo
  {author} {\bibfnamefont {Z.}~\bibnamefont {Zhong}}, \bibinfo {author}
  {\bibfnamefont {J.}~\bibnamefont {Sun}}, \bibinfo {author} {\bibfnamefont
  {P.}~\bibnamefont {Erhart}}, \bibinfo {author} {\bibfnamefont
  {Y.}~\bibnamefont {Su}},\ and\ \bibinfo {author} {\bibfnamefont
  {T.}~\bibnamefont {Ala-Nissila}},\ }\href {https://doi.org/10.1063/5.0106617}
  {\bibfield  {journal} {\bibinfo  {journal} {Journal of Chemical Physics}\
  }\textbf {\bibinfo {volume} {157}},\ \bibinfo {pages} {114801} (\bibinfo
  {year} {2022})}\BibitemShut {NoStop}%
\bibitem [{\citenamefont {Karabin}\ and\ \citenamefont
  {Perez}(2020)}]{KarPer20}%
  \BibitemOpen
  \bibfield  {author} {\bibinfo {author} {\bibfnamefont {M.}~\bibnamefont
  {Karabin}}\ and\ \bibinfo {author} {\bibfnamefont {D.}~\bibnamefont
  {Perez}},\ }\href {https://doi.org/10.1063/5.0013059} {\bibfield  {journal}
  {\bibinfo  {journal} {The Journal of Chemical Physics}\ }\textbf {\bibinfo
  {volume} {153}},\ \bibinfo {pages} {094110} (\bibinfo {year}
  {2020})}\BibitemShut {NoStop}%
\bibitem [{\citenamefont {Fan}\ \emph {et~al.}(2021)\citenamefont {Fan},
  \citenamefont {Zeng}, \citenamefont {Zhang}, \citenamefont {Wang},
  \citenamefont {Song}, \citenamefont {Dong}, \citenamefont {Chen},\ and\
  \citenamefont {Ala-Nissila}}]{FanZenZha21}%
  \BibitemOpen
  \bibfield  {author} {\bibinfo {author} {\bibfnamefont {Z.}~\bibnamefont
  {Fan}}, \bibinfo {author} {\bibfnamefont {Z.}~\bibnamefont {Zeng}}, \bibinfo
  {author} {\bibfnamefont {C.}~\bibnamefont {Zhang}}, \bibinfo {author}
  {\bibfnamefont {Y.}~\bibnamefont {Wang}}, \bibinfo {author} {\bibfnamefont
  {K.}~\bibnamefont {Song}}, \bibinfo {author} {\bibfnamefont {H.}~\bibnamefont
  {Dong}}, \bibinfo {author} {\bibfnamefont {Y.}~\bibnamefont {Chen}},\ and\
  \bibinfo {author} {\bibfnamefont {T.}~\bibnamefont {Ala-Nissila}},\ }\href
  {https://doi.org/10.1103/PhysRevB.104.104309} {\bibfield  {journal} {\bibinfo
   {journal} {Physical Review B}\ }\textbf {\bibinfo {volume} {104}},\ \bibinfo
  {pages} {104309} (\bibinfo {year} {2021})}\BibitemShut {NoStop}%
\bibitem [{\citenamefont {Drautz}(2019)}]{Dra19}%
  \BibitemOpen
  \bibfield  {author} {\bibinfo {author} {\bibfnamefont {R.}~\bibnamefont
  {Drautz}},\ }\href {https://doi.org/10.1103/PhysRevB.99.014104} {\bibfield
  {journal} {\bibinfo  {journal} {Physical Review B}\ }\textbf {\bibinfo
  {volume} {99}},\ \bibinfo {pages} {014104} (\bibinfo {year}
  {2019})}\BibitemShut {NoStop}%
\bibitem [{\citenamefont {Wierstra}\ \emph {et~al.}(2014)\citenamefont
  {Wierstra}, \citenamefont {Schaul}, \citenamefont {Glasmachers},
  \citenamefont {Sun}, \citenamefont {Peters},\ and\ \citenamefont
  {Schmidhuber}}]{WieSchGla14}%
  \BibitemOpen
  \bibfield  {author} {\bibinfo {author} {\bibfnamefont {D.}~\bibnamefont
  {Wierstra}}, \bibinfo {author} {\bibfnamefont {T.}~\bibnamefont {Schaul}},
  \bibinfo {author} {\bibfnamefont {T.}~\bibnamefont {Glasmachers}}, \bibinfo
  {author} {\bibfnamefont {Y.}~\bibnamefont {Sun}}, \bibinfo {author}
  {\bibfnamefont {J.}~\bibnamefont {Peters}},\ and\ \bibinfo {author}
  {\bibfnamefont {J.}~\bibnamefont {Schmidhuber}},\ }\href
  {http://jmlr.org/papers/v15/wierstra14a.html} {\bibfield  {journal} {\bibinfo
   {journal} {Journal of Machine Learning Research}\ }\textbf {\bibinfo
  {volume} {15}},\ \bibinfo {pages} {949} (\bibinfo {year} {2014})}\BibitemShut
  {NoStop}%
\bibitem [{\citenamefont {Eriksson}\ \emph {et~al.}(2019)\citenamefont
  {Eriksson}, \citenamefont {Fransson},\ and\ \citenamefont
  {Erhart}}]{EriFraErh19}%
  \BibitemOpen
  \bibfield  {author} {\bibinfo {author} {\bibfnamefont {F.}~\bibnamefont
  {Eriksson}}, \bibinfo {author} {\bibfnamefont {E.}~\bibnamefont {Fransson}},\
  and\ \bibinfo {author} {\bibfnamefont {P.}~\bibnamefont {Erhart}},\ }\href
  {https://doi.org/10.1002/adts.201800184} {\bibfield  {journal} {\bibinfo
  {journal} {Advanced Theory and Simulations}\ }\textbf {\bibinfo {volume}
  {2}},\ \bibinfo {pages} {1800184} (\bibinfo {year} {2019})}\BibitemShut
  {NoStop}%
\bibitem [{cal(2022)}]{calorine}%
  \BibitemOpen
  \href@noop {} {\bibinfo {title} {\textsc{calorine}}},\ \bibinfo
  {howpublished} {\url{https://calorine.materialsmodeling.org/}} (\bibinfo
  {year} {2022}),\ \bibinfo {note} {accessed: 2022-12-18}\BibitemShut {NoStop}%
\bibitem [{\citenamefont {Bussi}\ \emph {et~al.}(2007)\citenamefont {Bussi},
  \citenamefont {Donadio},\ and\ \citenamefont {Parrinello}}]{BusDonPar07}%
  \BibitemOpen
  \bibfield  {author} {\bibinfo {author} {\bibfnamefont {G.}~\bibnamefont
  {Bussi}}, \bibinfo {author} {\bibfnamefont {D.}~\bibnamefont {Donadio}},\
  and\ \bibinfo {author} {\bibfnamefont {M.}~\bibnamefont {Parrinello}},\
  }\href {https://doi.org/10.1063/1.2408420} {\bibfield  {journal} {\bibinfo
  {journal} {Journal of Chemical Physics}\ }\textbf {\bibinfo {volume} {126}},\
  \bibinfo {pages} {014101} (\bibinfo {year} {2007})}\BibitemShut {NoStop}%
\bibitem [{\citenamefont {Bernetti}\ and\ \citenamefont
  {Bussi}(2020)}]{BerBus20}%
  \BibitemOpen
  \bibfield  {author} {\bibinfo {author} {\bibfnamefont {M.}~\bibnamefont
  {Bernetti}}\ and\ \bibinfo {author} {\bibfnamefont {G.}~\bibnamefont
  {Bussi}},\ }\href {https://doi.org/10.1063/5.0020514} {\bibfield  {journal}
  {\bibinfo  {journal} {Journal of Chemical Physics}\ }\textbf {\bibinfo
  {volume} {153}},\ \bibinfo {pages} {114107} (\bibinfo {year}
  {2020})}\BibitemShut {NoStop}%
\bibitem [{\citenamefont {Frenkel}\ and\ \citenamefont
  {Ladd}(1984)}]{FreLad84}%
  \BibitemOpen
  \bibfield  {author} {\bibinfo {author} {\bibfnamefont {D.}~\bibnamefont
  {Frenkel}}\ and\ \bibinfo {author} {\bibfnamefont {A.~J.~C.}\ \bibnamefont
  {Ladd}},\ }\href {https://doi.org/10.1063/1.448024} {\bibfield  {journal}
  {\bibinfo  {journal} {The Journal of Chemical Physics}\ }\textbf {\bibinfo
  {volume} {81}},\ \bibinfo {pages} {3188} (\bibinfo {year}
  {1984})}\BibitemShut {NoStop}%
\bibitem [{\citenamefont {Frenkel}\ and\ \citenamefont
  {Smit}(2001)}]{FreSmi01}%
  \BibitemOpen
  \bibfield  {author} {\bibinfo {author} {\bibfnamefont {D.}~\bibnamefont
  {Frenkel}}\ and\ \bibinfo {author} {\bibfnamefont {B.}~\bibnamefont {Smit}},\
  }\href@noop {} {\emph {\bibinfo {title} {Understanding {{Molecular
  Simulation}}: {{From Algorithms}} to {{Applications}}}}}\ (\bibinfo
  {publisher} {{Academic Press}},\ \bibinfo {address} {San Diego},\ \bibinfo
  {year} {2001})\BibitemShut {NoStop}%
\bibitem [{\citenamefont {Freitas}\ \emph {et~al.}(2016)\citenamefont
  {Freitas}, \citenamefont {Asta},\ and\ \citenamefont {{de
  Koning}}}]{FreAstde16}%
  \BibitemOpen
  \bibfield  {author} {\bibinfo {author} {\bibfnamefont {R.}~\bibnamefont
  {Freitas}}, \bibinfo {author} {\bibfnamefont {M.}~\bibnamefont {Asta}},\ and\
  \bibinfo {author} {\bibfnamefont {M.}~\bibnamefont {{de Koning}}},\ }\href
  {https://doi.org/10.1016/j.commatsci.2015.10.050} {\bibfield  {journal}
  {\bibinfo  {journal} {Computational Materials Science}\ }\textbf {\bibinfo
  {volume} {112}},\ \bibinfo {pages} {333} (\bibinfo {year}
  {2016})}\BibitemShut {NoStop}%
\bibitem [{\citenamefont {Larsen}\ \emph {et~al.}(2017)\citenamefont {Larsen},
  \citenamefont {Mortensen}, \citenamefont {Blomqvist}, \citenamefont
  {Castelli}, \citenamefont {Christensen}, \citenamefont {Dułak},
  \citenamefont {Friis}, \citenamefont {Groves}, \citenamefont {Hammer},
  \citenamefont {Hargus}, \citenamefont {Hermes}, \citenamefont {Jennings},
  \citenamefont {Jensen}, \citenamefont {Kermode}, \citenamefont {Kitchin},
  \citenamefont {Kolsbjerg}, \citenamefont {Kubal}, \citenamefont {Kaasbjerg},
  \citenamefont {Lysgaard}, \citenamefont {Maronsson}, \citenamefont {Maxson},
  \citenamefont {Olsen}, \citenamefont {Pastewka}, \citenamefont {Peterson},
  \citenamefont {Rostgaard}, \citenamefont {Schiøtz}, \citenamefont {Schütt},
  \citenamefont {Strange}, \citenamefont {Thygesen}, \citenamefont {Vegge},
  \citenamefont {Vilhelmsen}, \citenamefont {Walter}, \citenamefont {Zeng},\
  and\ \citenamefont {Jacobsen}}]{LarMorBlo17}%
  \BibitemOpen
  \bibfield  {author} {\bibinfo {author} {\bibfnamefont {A.~H.}\ \bibnamefont
  {Larsen}}, \bibinfo {author} {\bibfnamefont {J.~J.}\ \bibnamefont
  {Mortensen}}, \bibinfo {author} {\bibfnamefont {J.}~\bibnamefont
  {Blomqvist}}, \bibinfo {author} {\bibfnamefont {I.~E.}\ \bibnamefont
  {Castelli}}, \bibinfo {author} {\bibfnamefont {R.}~\bibnamefont
  {Christensen}}, \bibinfo {author} {\bibfnamefont {M.}~\bibnamefont {Dułak}},
  \bibinfo {author} {\bibfnamefont {J.}~\bibnamefont {Friis}}, \bibinfo
  {author} {\bibfnamefont {M.~N.}\ \bibnamefont {Groves}}, \bibinfo {author}
  {\bibfnamefont {B.}~\bibnamefont {Hammer}}, \bibinfo {author} {\bibfnamefont
  {C.}~\bibnamefont {Hargus}}, \bibinfo {author} {\bibfnamefont {E.~D.}\
  \bibnamefont {Hermes}}, \bibinfo {author} {\bibfnamefont {P.~C.}\
  \bibnamefont {Jennings}}, \bibinfo {author} {\bibfnamefont {P.~B.}\
  \bibnamefont {Jensen}}, \bibinfo {author} {\bibfnamefont {J.}~\bibnamefont
  {Kermode}}, \bibinfo {author} {\bibfnamefont {J.~R.}\ \bibnamefont
  {Kitchin}}, \bibinfo {author} {\bibfnamefont {E.~L.}\ \bibnamefont
  {Kolsbjerg}}, \bibinfo {author} {\bibfnamefont {J.}~\bibnamefont {Kubal}},
  \bibinfo {author} {\bibfnamefont {K.}~\bibnamefont {Kaasbjerg}}, \bibinfo
  {author} {\bibfnamefont {S.}~\bibnamefont {Lysgaard}}, \bibinfo {author}
  {\bibfnamefont {J.~B.}\ \bibnamefont {Maronsson}}, \bibinfo {author}
  {\bibfnamefont {T.}~\bibnamefont {Maxson}}, \bibinfo {author} {\bibfnamefont
  {T.}~\bibnamefont {Olsen}}, \bibinfo {author} {\bibfnamefont
  {L.}~\bibnamefont {Pastewka}}, \bibinfo {author} {\bibfnamefont
  {A.}~\bibnamefont {Peterson}}, \bibinfo {author} {\bibfnamefont
  {C.}~\bibnamefont {Rostgaard}}, \bibinfo {author} {\bibfnamefont
  {J.}~\bibnamefont {Schiøtz}}, \bibinfo {author} {\bibfnamefont
  {O.}~\bibnamefont {Schütt}}, \bibinfo {author} {\bibfnamefont
  {M.}~\bibnamefont {Strange}}, \bibinfo {author} {\bibfnamefont {K.~S.}\
  \bibnamefont {Thygesen}}, \bibinfo {author} {\bibfnamefont {T.}~\bibnamefont
  {Vegge}}, \bibinfo {author} {\bibfnamefont {L.}~\bibnamefont {Vilhelmsen}},
  \bibinfo {author} {\bibfnamefont {M.}~\bibnamefont {Walter}}, \bibinfo
  {author} {\bibfnamefont {Z.}~\bibnamefont {Zeng}},\ and\ \bibinfo {author}
  {\bibfnamefont {K.~W.}\ \bibnamefont {Jacobsen}},\ }\href
  {https://doi.org/10.1088/1361-648X/aa680e} {\bibfield  {journal} {\bibinfo
  {journal} {Journal of Physics: Condensed Matter}\ }\textbf {\bibinfo {volume}
  {29}},\ \bibinfo {pages} {273002} (\bibinfo {year} {2017})}\BibitemShut
  {NoStop}%
\bibitem [{\citenamefont {Bl\"ochl}(1994)}]{Blo94}%
  \BibitemOpen
  \bibfield  {author} {\bibinfo {author} {\bibfnamefont {P.~E.}\ \bibnamefont
  {Bl\"ochl}},\ }\href {https://doi.org/10.1103/PhysRevB.50.17953} {\bibfield
  {journal} {\bibinfo  {journal} {Physical Review B}\ }\textbf {\bibinfo
  {volume} {50}},\ \bibinfo {pages} {17953} (\bibinfo {year}
  {1994})}\BibitemShut {NoStop}%
\bibitem [{\citenamefont {Kresse}\ and\ \citenamefont
  {Hafner}(1993)}]{KreHaf93}%
  \BibitemOpen
  \bibfield  {author} {\bibinfo {author} {\bibfnamefont {G.}~\bibnamefont
  {Kresse}}\ and\ \bibinfo {author} {\bibfnamefont {J.}~\bibnamefont
  {Hafner}},\ }\href {https://doi.org/10.1103/PhysRevB.47.558} {\bibfield
  {journal} {\bibinfo  {journal} {Physical Review B}\ }\textbf {\bibinfo
  {volume} {47}},\ \bibinfo {pages} {558} (\bibinfo {year} {1993})}\BibitemShut
  {NoStop}%
\bibitem [{\citenamefont {Kresse}\ and\ \citenamefont
  {Furthm\"uller}(1996)}]{KreFur96}%
  \BibitemOpen
  \bibfield  {author} {\bibinfo {author} {\bibfnamefont {G.}~\bibnamefont
  {Kresse}}\ and\ \bibinfo {author} {\bibfnamefont {J.}~\bibnamefont
  {Furthm\"uller}},\ }\href {https://doi.org/10.1016/0927-0256(96)00008-0}
  {\bibfield  {journal} {\bibinfo  {journal} {Computational Materials Science}\
  }\textbf {\bibinfo {volume} {6}},\ \bibinfo {pages} {15} (\bibinfo {year}
  {1996})}\BibitemShut {NoStop}%
\end{thebibliography}
\end{document}